\documentclass[]{aastex63}
\usepackage{graphicx,times}
\usepackage{amssymb,amsmath}
\bibpunct{(}{)}{;}{a}{}{,}
\usepackage{CJK}
\usepackage[graphicx]{realboxes}

\usepackage{multirow}
\usepackage{rotating}

\newcommand{\s}{s$^{-1}$}

\newcommand{\Ms}{$M_{\odot}$}
\graphicspath{{./}{fig/}}

\shorttitle{UV Fields in Gould's Belt}
\shortauthors{Xia et al.}

\begin{document}

\title{The Distribution of UV Radiation Field in the Molecular Clouds of Gould Belt}

\correspondingauthor{Di Li, Ningyu Tang, Qijun Zhi}
\email{dili@nao.cas.cn,nytang@ahnu.edu.cn,qjzhi@gznu.edu.cn}

\author[0000-0003-3726-570X]{Jifeng Xia}   
\affiliation{School of Physics and Electronic Science, Guizhou Normal University, Guiyang, 550025, China}
\affiliation{National Astronomical Observatories, Chinese Academy of Sciences, Beijing 100101, China}
\affiliation{University of Chinese Academy of Sciences, Beijing 100049, China}

\author[0000-0002-2169-0472]{Ningyu Tang}
\affiliation{Department of Physics, Anhui Normal University, Wuhu, Anhui 241002, China}

\author{Qijun Zhi}
\affiliation{School of Physics and Electronic Science, Guizhou Normal University, Guiyang, 550025, China}
\affiliation{Guizhou Provincial Key Laboratory of Radio Astronomy and Data Processing, Guizhou Normal University, Guiyang 550001, China}

\author[0000-0002-9151-1388]{Sihan Jiao}
\affiliation{National Astronomical Observatories, Chinese Academy of Sciences, Beijing 100101, China}
\affiliation{University of Chinese Academy of Sciences, Beijing 100049, China}

\author[0000-0002-2738-146X]{Jinjin Xie}
\affiliation{Shanghai Astronomical Observatory, CAS, Shanghai,200030, China}

\author{Gary~A. Fuller}
\affiliation{Jodrell Bank Centre for Astrophysics, Department of Physics \& Astronomy, The University of Manchester, Manchester M13 9PL, United Kingdom}
\affiliation{I. Physikalisches Institut, University of Cologne, Z\"ulpicher Str. 77, 50937 K\"oln, Germany}

\author{Paul F. Goldsmith}
\affiliation{Jet Propulsion Laboratory, California Institute of Technology, 4800 Oak Grove Drive, Pasadena, CA 91109, USA}

\author[0000-0003-3010-7661]{Di Li}
\affiliation{National Astronomical Observatories, Chinese Academy of Sciences, Beijing 100101, China}
\affiliation{University of Chinese Academy of Sciences, Beijing 100049, China}

\begin{abstract}

The distribution of ultraviolet (UV) radiation field provides critical constraints on the physical environments of molecular clouds. 
Within 1\,kpc of our solar system and fostering protostars of different masses, the giant molecular clouds in the Gould Belt present an excellent opportunity to resolve the UV field structure in star forming regions. We performed spectral energy distribution (SED) fitting of the archival data from the {\it Herschel} Gould Belt Survey (HGBS). Dust radiative transfer analysis with the DUSTY code were applied to 23 regions in 14 molecular complexes of the Gould Belt, resulting in the spatial distribution of radiation field in these regions. 
For 10 of 15 regions with independent measurements of star formation rate, their star formation rate and UV radiation intensity largely conform to a linear correlation found in previous studies.  

\end{abstract}


\section{Introduction}           
\label{sect:intro}

Interstellar medium (ISM) is the cradle of star formation. Its evolution is strongly affected by the ultra-violet (UV) radiation. By ejecting electrons from dust grains and directly exciting atoms and molecules, UV photons ionize atoms, dissociate molecules and heat gases \citep[e.g.,][]{1985ApJ...291..722T, 2001RvMP...73.1031F, 2011piim.book.....D}.  \citet{1998ARA&A..36..189K} found a tight correlation between UV radiation and the star formation rate in galaxies. In Galactic studies, most previous studies focus on the distribution of UV radiation field of individual nearby regions \citep[e.g. ][]{1999A&A...344..342L,2010A&A...521L..19P}. 

Within 1\,kpc of our Solar system, the Gould Belt containing a lot of molecular complexes provides an excellent opportunity to investigate the relationship between UV intensity and star formation rate under diverse environments \citep{2007PASP..119..855W}. Most molecular complexes of the Gould Belt harbor bright young OB stellar clusters or star-forming regions while the rest show little sign of star formation. 
For instance, Orion complex is a widely studied giant molecular region with abundant clustering OB stars  and turbulent massive star formation \citep[e.g.,][]{1991ApJ...368..432L,1998ApJS..118..517T}. Similar to Orion, Serpens/Aquila Rift is a rich complex with well-known massive star-forming regions, e.g., W40 HII region that contains embedded young high-mass stars \citep[e.g.,][]{1999PASJ...51..851K}. Although the Cepheus region contains an OB star (HD 200775 in Cep 1172), it has been generally considered as a low to intermediate mass star-forming region.  

Lupus dark-cloud complex was found to be surrounded by about 70 T Tauri stars, with no indication of massive OB star inside \citep[e.g.,][]{1999PASJ...51..895H}. The Chamaeleon-Musca dark-cloud complex including Cha I, II, III, and the Musca dark lane, is a region with low-mass star formation \citep[e.g.,][]{1999A&A...345..965C, 2001PASJ...53.1071M}.  IC 5146 is a filamentary dark cloud with scattered low star formation activity \citep[e.g.,][]{1994ApJS...95..419D}. Pipe Nebula has been a primary example of little signs of disturbance from star formation \citep[e.g.,][]{1999PASJ...51..871O}. No star formation was found in the  Polaris flare, a high-latitude translucent cloud \citep[e.g.,][]{1990ApJ...353L..49H}.

Utilizing the Spectral and Photometric Imaging Receiver (SPIRE) and Photodetector Array Camera and Spectrometer (PACS) instruments on board the {\it Herschel} \footnote{{\it Herschel} is an European Space Agency (ESA) space observatory with science instruments provided by European-led Principal Investigator consortia and with important participation from NASA.} Space Observatory, the {\it Herschel} Gould Belt Survey (HGBS)  covered a substantial fraction of the Gould Belt. Specifically, images were taken at 250, 350, and 500 $\mu$m for regions with $A\rm_V > $ 3 mag  with SPIRE and at 70 and 160 $\mu$m for those with $A\rm_V > $ 6 mag with  PACS. This survey covers the following 14 molecular complexes:  Aquila, Cepheus, Chamaleon, Corona Australis, IC 5146, Lupus, Musca, $\rho$ Oph, Orion, Perseus, Pipe nebula, Polaris, Taurus, and Serpens.

The dust radiative transfer model, DUSTY \citep{1997MNRAS.287..799I, 2000ASPC..196...77N} accommodate  different kinds of geometry and radiation parameters. \cite{2003ApJ...587..262L} demonstrated the utilities of the DUSTY code to derive the UV intensity of the Orion clouds based on fitting the dust temperature data. We further develop the recipe and apply it to the full set of the HGBS data. 

This paper is organized as follows. In Section \ref{sec:data}, we introduce the data information of column density, dust temperature  and star formation rate toward the HGBS molecular complexes. The DUSTY radiative transfer code and method for calculating the UV radiation intensity map are described in Section \ref{sec:method}. Results and further discussions are presented in Section  \ref{sec:results} and \ref{sec:discussion}, respectively.  The summary is in Section \ref{sec:summary}.

\section{Data}
\label{sec:data}

\subsection{{\it Herschel} Dust Continuum Emission}\label{subsection:dust}

Dust emission is almost always optically thin at (sub)millimete wavelengths and can thus act as a surrogate tracer of the total (gas + dust) mass along the line of sight (LOS) \citep[][]{2014A&A...562A.138R}. 
The HGBS took a census in the nearby (0.5\,kpc) molecular cloud complexes for an extensive imaging survey of the densest portions of the Gould Belt, down to a 5$\sigma$ column sensitivity $N_{H_{2}}$ $\sim$ 10$^{21}$\,cm$^{-2}$ or $A_{V}$ $\sim$ 1 mag \citep[][]{2010A&A...518L.102A}.
We use the 

HGBS data to generate the column density map of 23 molecular clouds that belongs to 14 molecular complexes of the Gould Belt . 

 The molecular clouds studied in this paper are Aquaila M2 \citep{2010A&A...518L.102A} , Cep 1151, Cep 1172, Cep 1228, Cep 1241, Cep 1251 \citep{2020ApJ...904..172D}, Cham I, Cham II, Cham III \citep{2012A&A...545A.145W}, \citep{2014A&A...568A..98A}, CraNS \citep{2018A&A...615A.125B}, IC 5146 \citep{2019A&A...621A..42A}, Lup I, Lup III, Lup IV \citep{2013A&A...549L...1R}, Musca \citep{2016A&A...590A.110C}, $\rho$ Oph  \cite{2014A&A...562A.138R}, Orion B \citep{2020A&A...635A..34K}, Orion A(Jiao et al. accepted by SCPMA), Perseus \citep{2012A&A...547A..54P}, Pipe \citep{2012A&A...541A..63P}, Polaris \citep{2010A&A...518L.102A}, Taurus \citep{2019A&A...621A..42A}, and Serpens\citep{2021MNRAS.500.4257F}.

\subsection{Deriving Dust Temperature and Column Density Based on SED Fitting}
\label{subsection:sed}

The dust temperature and column density map toward 22 of 23 regions were downloaded from HGBS Archive\footnote{The website of HGBS Archive: http://www.herschel.fr/cea/gouldbelt/en/index.php}. We obtained dust distribution and improved the image quality of the Orion A region, based on a novel image combination technique (Jiao et al.\ 2022, accepted). \footnote{The column density and dust temperature of Orion A was not published on the website of HGBS archive by 2022.} The procedure to derive the dust temperature and dust/gas column density images of Orion A is similar to that in \cite{2014A&A...562A.138R}. Before performing any SED fitting, all images at multiple bands were convolved into beam size of 36$''$.3 at 500 $\mu$m.
We weighted the data points by the measured noise level in the least-squares fits. 
We adopted the dust opacity per unit mass at 300 $\mu$m of 0.1 cm$^{2}$ g$^{-2}$ \citep{1983QJRAS..24..267H}, and assumed a gas-to-dust mass ratio of 100.
As modified black-body assumption, the flux density $S_{\nu}$ at a certain observing frequency $\nu$ is given by
\begin{equation}
S_{\nu} = \Omega_{m}B_{\nu}(T_{d})(1-e^{-\tau_{\nu}}),\label{eq:Snu}
\end{equation} 
where $B_{\nu}(T_{d})$ is the Planck function at temperature $T_{d}$,  $\Omega_{m}$ is the beam size. The total column density $N$ of gas and dust can be approximated by
\begin{equation}
N=\frac{\tau_{\nu}}{\kappa_{\nu}\mu m_{H}},\label{eq:N}
\end{equation}
where the dust opacity $\kappa_{\lambda}=\kappa_{\mbox{\tiny{300$\mu$m}}}(\lambda/300\,\mu m)^{\beta}$ ($\beta$ was fixed to a constant value of 1.8),  $\mu$ = 2.8 is the mean molecular weight, $m_{H}$ is the mass of a hydrogen atom. The grey-body dust temperature ($T_d$) thus calculated has ignored the dependence of dust temperature on grain size \citep{1999ApJ...522..897L}, but has been shown to be within a couple of kelvins of the gas temperatures in well coupled regions \citep{2001ApJ...557..736G,2013ApJ...768L...5L,2020MNRAS.499.4432W, 2021SCPMA..6479511X}.
The effect of scattering opacity \citep[e.g.,][]{2019ApJ...877L..22L} can be safely ignored in the case given that focusing on $>$0.05\,pc scale structures.

\subsection{Star Formation Rates}\label{subsection:sfr}

The SFRs were determined through $\mbox{SFR}$ = $N(\mbox{\scriptsize YSO}) \langle M \rangle \tau^{-1}$, in which $N(\mbox{\scriptsize YSO})$ is the YSOs number in the molecular region, $\langle M \rangle$ is the mean mass of stars, and $\tau$ is the relevant evolution timescale.  

The identification of YSOs requires careful discrimination against background stars and galaxies. 
Star-forming galaxies were the most problematic source of contaminants.
A combination of color - color and color - magnitude diagrams of both Spitzer and 2MASS data were adopted to reject contaminants.
The detailed descriptions of this process are shown in \citet{2009ApJS..181..321E}.
\cite{2013AJ....145...94D} provides a YSO catalog for Gould Belt clouds.
To convert number of YSOs to mass of forming stars, \cite{2014ApJ...782..114E} adopted mean mass of stars M$_\star$ = 0.5 M{$_ \odot $}, based on a fully sampled initial mass function \citep[][]{2002Sci...295...82K}. The relevant timescales were derived by classifying YSOs into standard SED classes based on 2 to 24 $\mu$m data.

Table~\ref{tab:SF} lists the star formation rates \citep{2010ApJ...724..687L, 2014ApJ...782..114E} we adopted in this paper.  

\begin{table}[h!]
\centering
\begin{minipage}[]{100mm}
\caption[]{Star formation rate of 15 molecular regions.\label{tab:SF}}\end{minipage}
\begin{tabular}{lccccr}
 \hline\noalign{\smallskip}
Cloud Name & RA & DEC & Distance$^a$  &Star Formation Rate$^a$  & Category$^b$ \\
         & (deg)& (deg) & (pc) & (10$^{-6}$\Ms yr$^{-1}$)  &    \\
\hline
Orion A    & 84.52 & -7.03 & 423 & 715 & I \\
Orion B    & 86.90 & 0.10 & 423 & 159 & I \\
Serpens    &  278.78&  4.6E-06& 429& 56 & I \\
Perseus    & 53.92  & 31.53 & 250 & 150 & I \\
$\rho$ Oph  & 246.87 & -24.21 & 125 & 79 & I \\
Musca      & 186.81 & -71.54 & 200 & 3 & I \\
Lupus III  & 242.51 & -39.08 &200 &  17 & II \\
Lupus IV   & 241.15 & -42.07 &150 & 3  & II \\
Aquila     & 277.43 & -2.78& 260 & 322 & II \\
IC 5146    & 327.19 & 47.49 & 460 & 24 & II \\
Cham I     & 165.46 & -77.40 & 150 & 20.5 & II \\
Cham II    & 195.23  & -77.41 & 178 & 6 & II \\
Cham III   & 190.47 & -79.82 & 150 & 1 & II \\
Pipe       & 260.78 & -26.40 & 145 & 5 & II \\
Taurus    & 65.08  & 27.72 & 153 & 84 & II \\  
  \hline\noalign{\smallskip}
  \noalign{\smallskip}\hline
\end{tabular}
\tablecomments{The coordinates are determined by the data of HGBS.  \\
$^{a}$The values of distances and star formation rates are obtained from \citet{2010ApJ...724..687L} and \citet{2014ApJ...782..114E}. \\ 
$^{b}$The region with both OB stars and YSOs  are defined as Category I; the region with YSOs only are defined as Category II.}

\end{table}

\section{Method}
\label{sec:method}

\subsection{The DUSTY code}
\label{sec:dusty}

The public DUSTY code \citep{1997MNRAS.287..799I} solves the radiative transfer problems through a fully scale-free method. By adopting this scaling method, the DUSTY code solves the spherically symmetric (1-D) problem with a single central radiation source and surrounding spherically symmetric dusty envelope, in which the radial dust density profile is arbitrary. Besides, a dusty plane-parallel slab with illumination from one or both sides at an arbitrary angle is available too. 

This code utilizes the scaling properties to minimize the number of input parameters.  Parameters  describing the external radiation, dust and gas properties, cloud geometry are needed for inputs. For the case of spherical geometry, they include number, spectral shape and flux of the external source, chemical composition of dust,  the lower and upper limit of dust optical depth, and density distribution of dust.  For the case of slab geometry, an extra selection of source side is needed.

Once the input parameters are chosen, the DUSTY code outputs the value of dust temperature as a function of gas optical depth.

\subsection{Parameter Setting in $\it$ DUSTY Code}
\label{subsec:dusty-para}

We adopted the slab geometry for our calculations. The incident angle between UV radiation and the slab, $\theta$  is treated as 0 degree, $\theta$= 0$^\circ$ , indicating perpendicular UV radiation toward the slab. The selected radiation wavelength of 0.365\,$\mu$m locates in the central wavelength of UV band. The dust sublimation temperature, which is the highest temperature the dust grains can exist, is chosen to be the common value of 1500 K.  The value range of gas optical depth and dust temperature were chosen as [0,30] and [0, 50] K, separately.

With the above selections, the UV radiation intensity, G$_0$ is the only parameter that determines the relationship between dust temperature and gas optical depth. As for the optional UV radiation flux received by one side of the slab, six sets of UV intensity (G$_0$= 1, 10, 31.6, 100, 316, and 1000) compared to standard Habing field \citep[e.g.,][]{2017A&A...602A..49H} of 1.6 $\times$ 10$^{-6}$\,W/m$^2$ were introduced for calculation.

\subsection{Distribution of UV Intensity Field }
\label{sec:contour}

In order to produce the spatial distribution of UV radiation field of the Gould Belt, we combined the DUSTY model calculations and the HGBS data. 

The derived H$_2$ column density and dust temperature map of HGBS molecular complexes with pixel size of 3$^{\prime \prime}$ were convolved and re-sampled with the beam size of 36.3$^{\prime\prime}$ at 500 $\mu$m. We are focusing on dense molecular regions, the contribution from atomic hydrogen can be ignored. Thus the H$_2$ column density was converted into gas optical depth at V band through the following equation \citep{1978ApJ...224..132B, 1968nim..book..221G, 2009ApJS..180..125R}    

\begin{equation}
\tau_V = \frac{A_V}{1.086}=\frac{1.07\times 10^{-21}N_{H_2}}{1.086}. \label{eq:tauv}
\end{equation}

\begin{figure}[ht!]
\centering
 \includegraphics[height=0.60\textwidth,width=0.90\textwidth,trim={{0.05\textwidth} {0.05\textwidth}  {0.05\textwidth}  {0.05\textwidth} },clip]{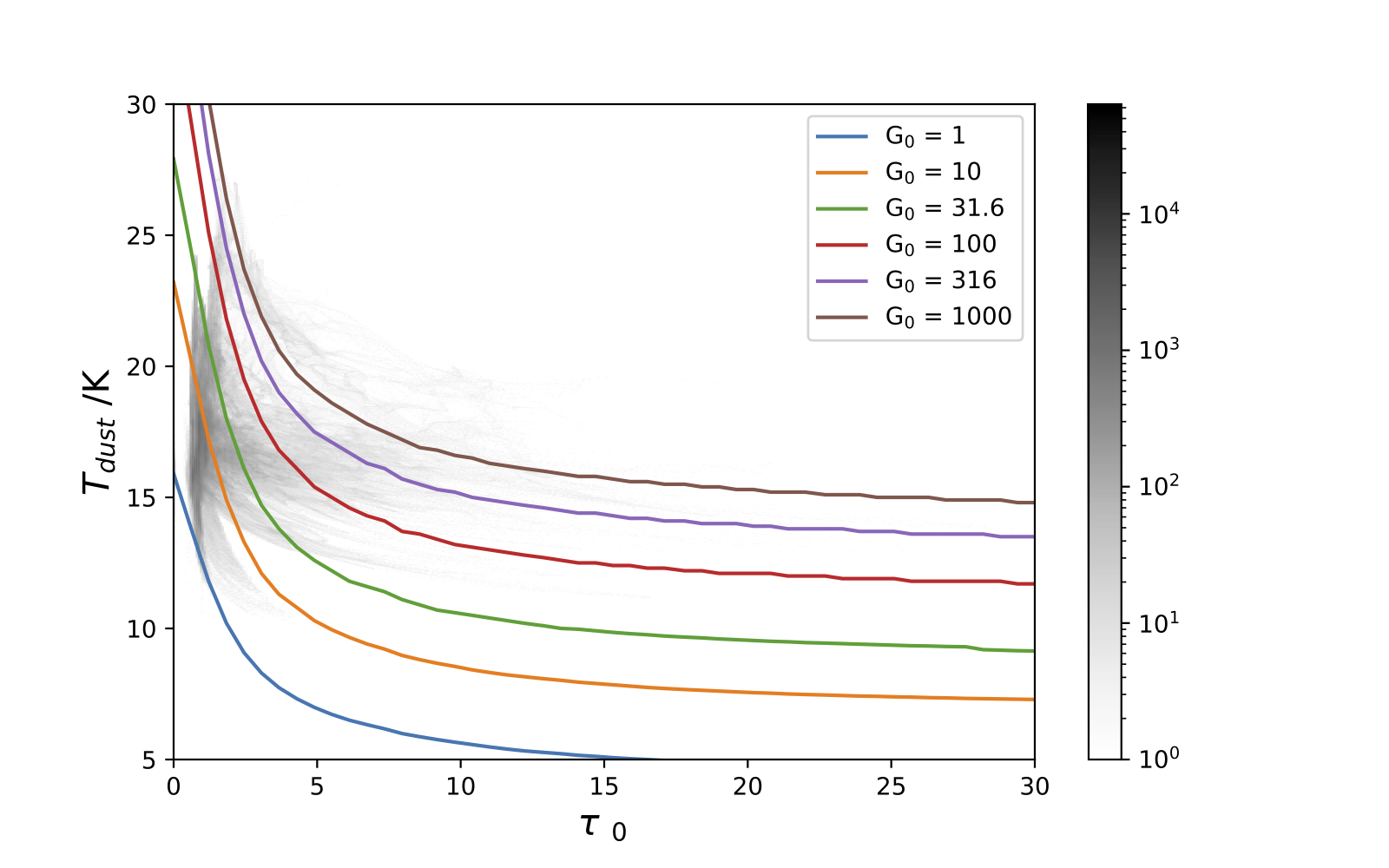}
 \caption{Relationship between dust temperature and optical depth.  Results for the observed $\rho$ Oph data are shown with grey point. Solid color lines represent model calculations from DUSTY code with different UV intensity value, G$_0$.  }
 
 \label{fig:tau-dust}
 \end{figure}

As an example, the derived relationship between T$_{dust}$ and gas optical depth ($\tau_V$) of all pixels for $\rho$ Oph cloud can be found in Fig. \ref{fig:tau-dust}. We obtained the UV intensity of each pixel by interpolating the results from observations with that from DUSTY model calculations. The spatial distribution of specific UV radiation field of each HGBS molecular complex can be derived. 

\section{Results}
\label{sec:results}

After applying the procedures described in section \ref{sec:method}, we  obtained the UV intensity distribution of 23 regions in 14 molecular complexes of the Gould Belt.

\subsection{Spatial Distribution of UV Intensity}

As an example, We presented spatial distribution of selected UV intensity (G$_0$= 1, 10, 31.6, 100, 316, and 1000)  for three  molecular complexes: Orion (Category I), Aquaila (Category II) and Polaris (Category III). Descriptions of the rest complexes are presented in Appendix.

\subsubsection{Orion}
Orion molecular complex is the most active star forming region within 500 pc \citep{2012AJ....144..192M}. Orion A and Orion B  molecular clouds are covered by HGBS. Locating  in Orion A, Orion Nebular cluster is a significant laboratory for understanding the initial mass function. The Orion B molecular cloud is one of the clouds scattered along the region named Orion-Eridanus superbubble, which was created by supernova explosions \citep{2020A&A...635A..34K}. The Orion B cloud is $\sim$ 423 pc away and covers an area of $\sim$ 6.8 $\times$ 8.6deg{$^2$} \citep{2014ApJ...782..114E}. The total mass of these two molecular clouds exceed 2 $\times$ 10{$^5$} M{$_{\odot}$} \citep{2012AJ....144..192M}.

\begin{figure}[h!]
 \centering
 \includegraphics[width=0.45\textwidth,height=0.45\textwidth,trim={{0.05\textwidth} {0.15\textwidth}  {0.05\textwidth}  {0.1\textwidth} },clip]{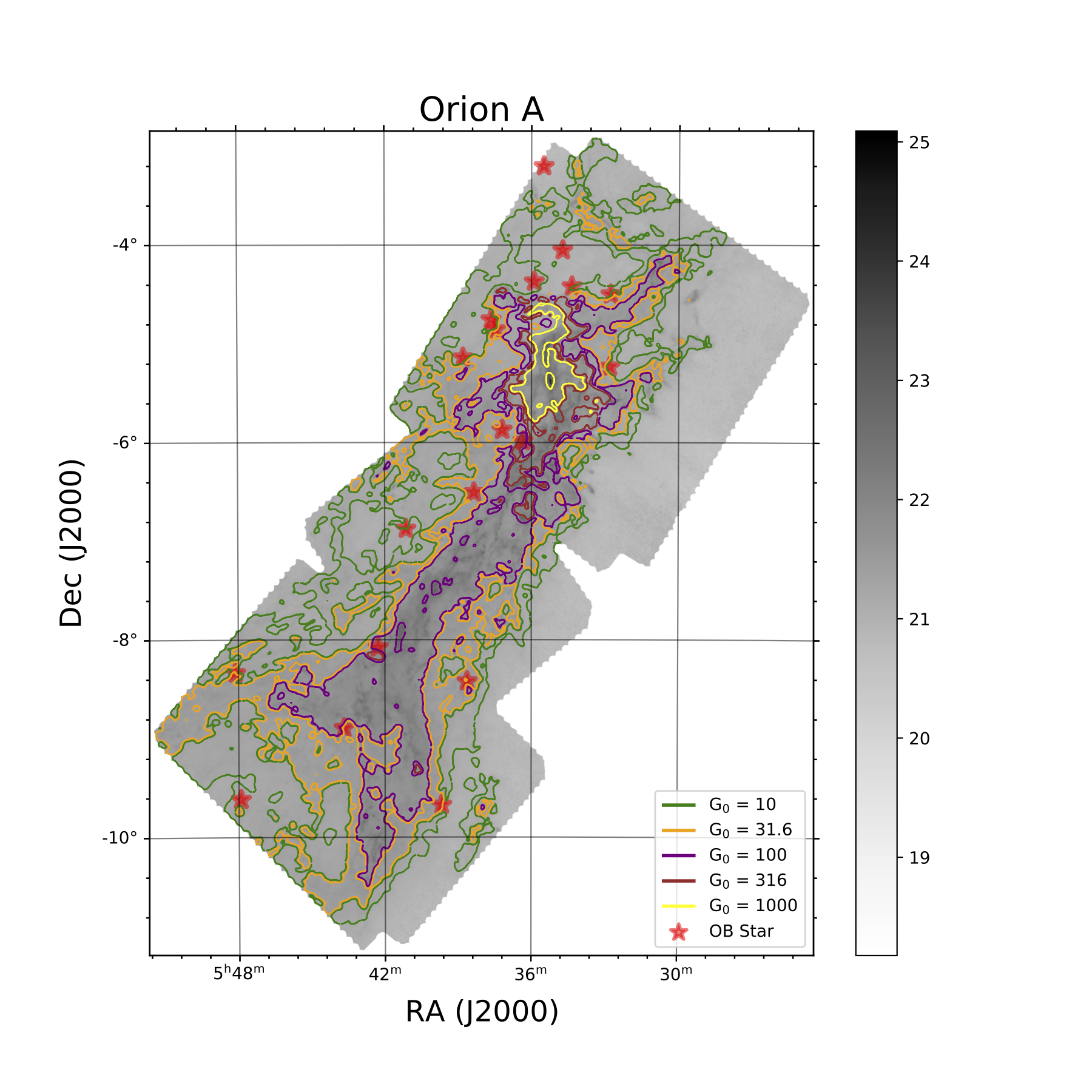}
  \includegraphics[width=0.45\textwidth,height=0.45\textwidth,trim={{0.01\textwidth} {0.15\textwidth}  {0.05\textwidth}  {0.1\textwidth} },clip]{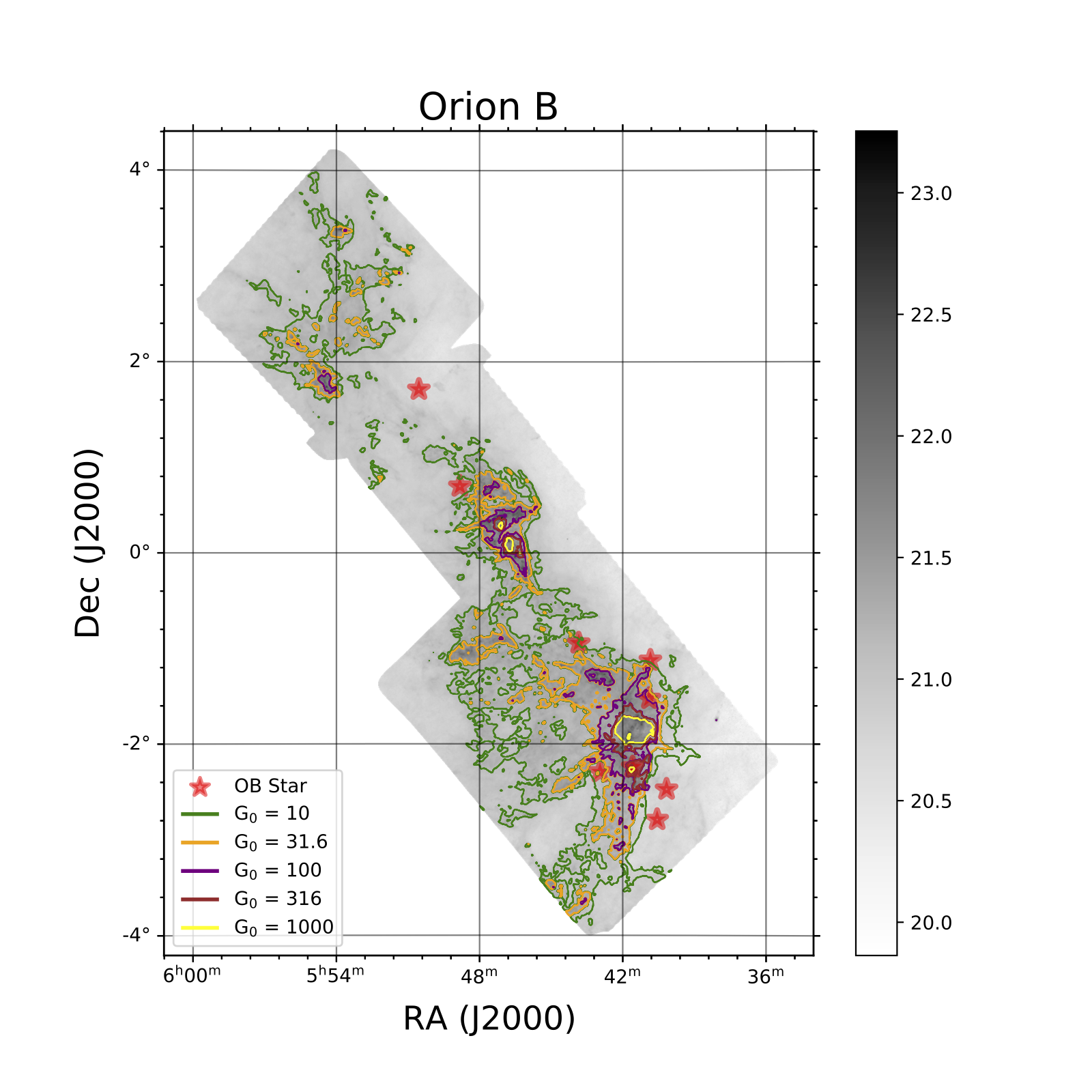}
 \caption{Contours of UV intensity overlaid on H$_2$ column density map of the Orion A ($left\ panel$) and Orion B ($right\ panel$) region. The column density N$(H_2)$ are shown with log$_{10} N(H_2)$ value. The red star implies the location of OB star in this region.} 
 \label{fig:orion}
 \end{figure}

As shown in Fig. \ref{fig:orion}, the UV intensity  of these two regions show tight correlation with the OB stars.

\subsubsection{Aquila}

The Aquila field is a very active star-forming region at a distance of about 260 pc \citep{2014ApJ...782..114E}. With size of $\sim$ 7.56 pc, the mass of this molecualr complex is about 24446 M$_{\odot}$, of which two-thirds are composed of dense cores \citep{2014ApJ...782..114E}. The existence of dense cloud leads to highest level of background cloud emission in HGBS. Cluster of YSOs but no OB stars were detected in this region. The number of YSOs in this region exceeds 1000. 

\begin{figure}[h!]
 \centering
 \includegraphics[width=0.80\textwidth,height=0.70\textwidth,trim={{0.05\textwidth} {0.3\textwidth}  {0.05\textwidth}  {0.1\textwidth} },clip]{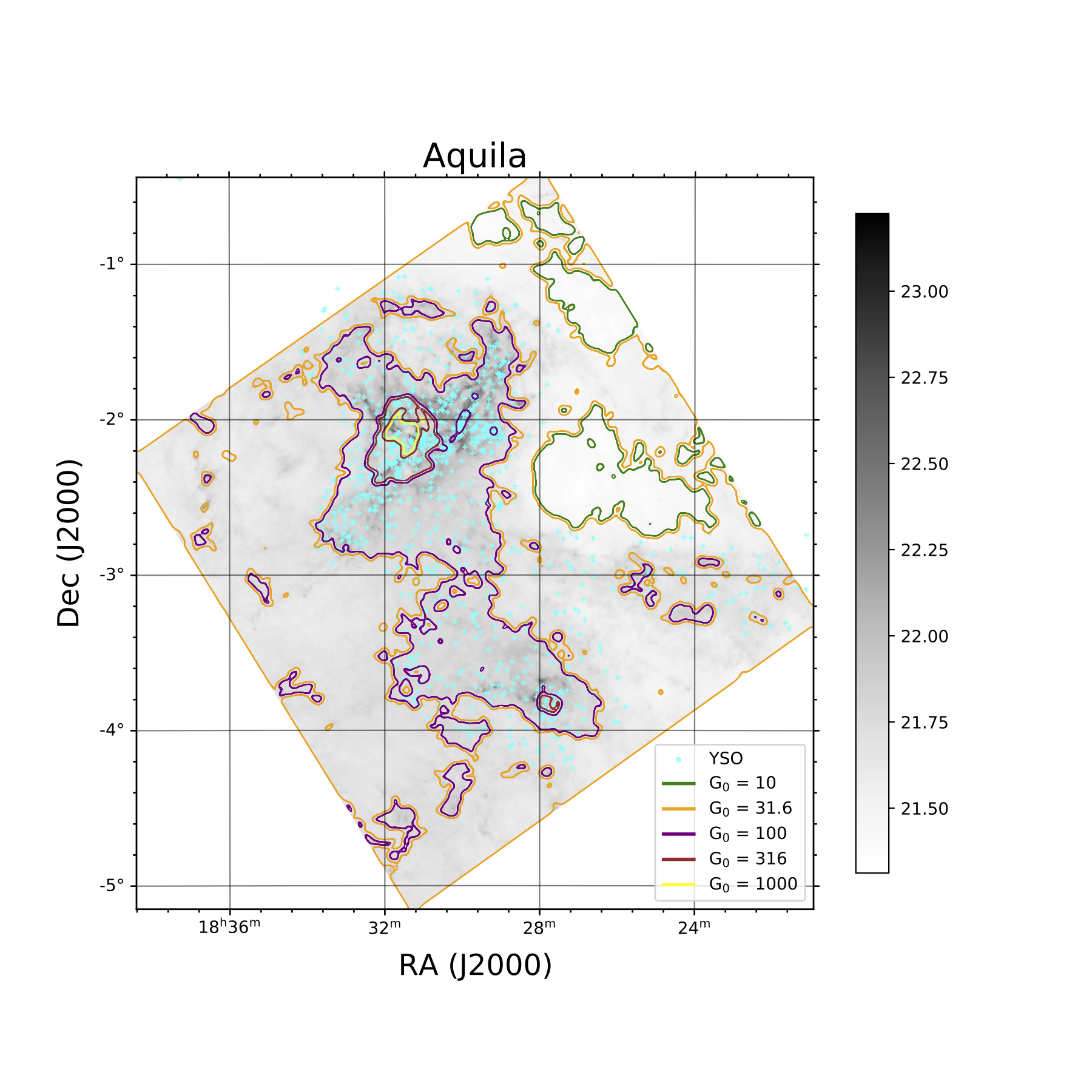}
\caption{Contours of UV intensity overlaid on H$_2$ column density map of the Aquila molecular complex. The column density N$(H_2)$ are shown with log$_{10} N(H_2)$ value. The cyan dots represent the locations of YSOs.}
\label{fig:aquila}
\end{figure}

As shown in Figure \ref{fig:aquila}, the intensity of the radiation field distribution varies from G$_0$=1 to G$_0$= 1000. We found a certain correlation between the spatial distribution of YSOs and UV radiation field. The possible reason responsible for this is discussed in section \ref{sec:discussion}.

\subsubsection{Polaris}
\begin{figure}[h!]
\centering
\includegraphics[width=0.80\textwidth,height=0.70\textwidth,trim={{0.15\textwidth} {0.45\textwidth}  {0.05\textwidth}  {0.1\textwidth} },clip]{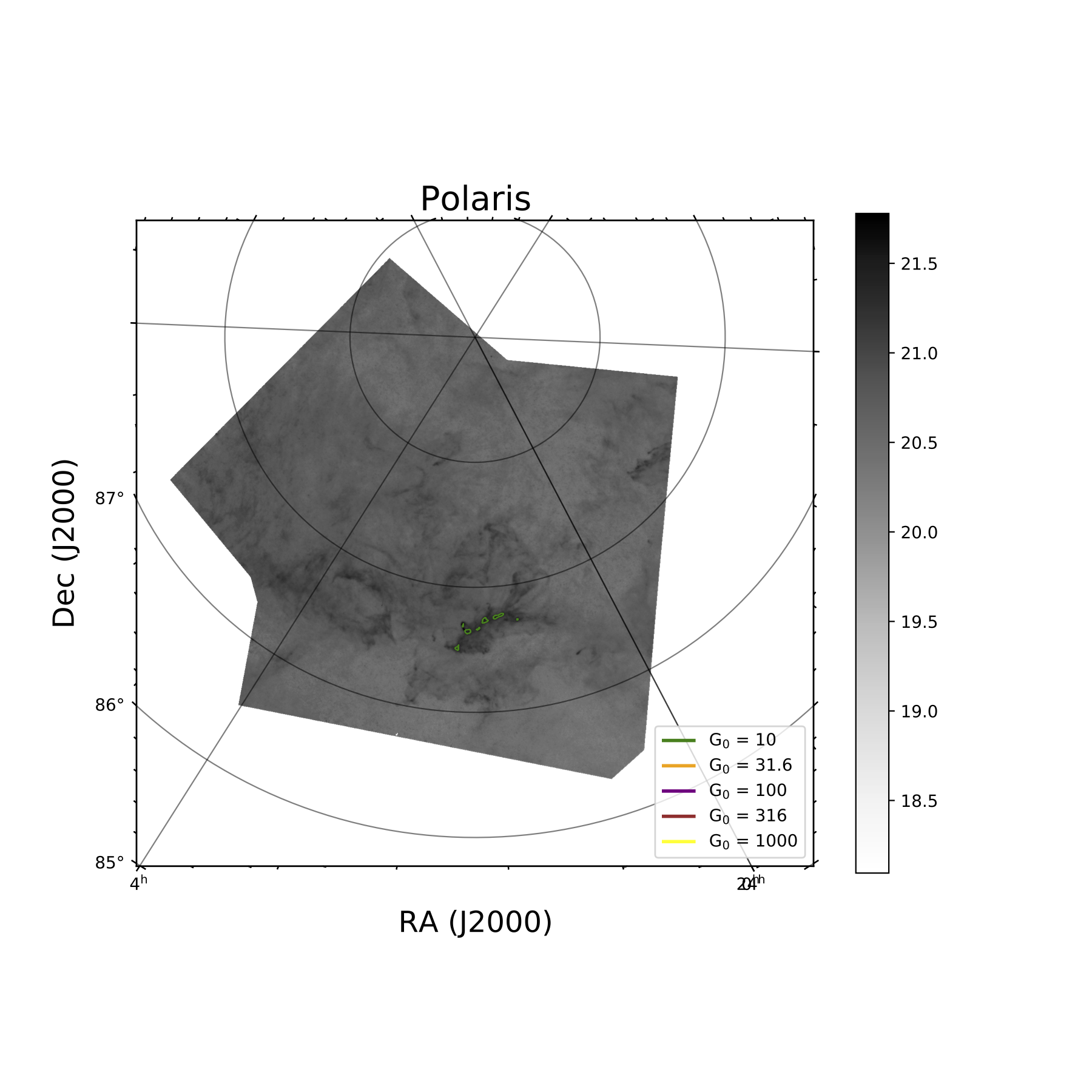}
\caption{Contours of UV intensity overlaid on H$_2$ column density map of the Polaris region. The labels of the figure are the same as that in Fig. \ref{fig:aquila}. No YSOs or OB star were found in this region.} 
 
\label{fig:polaris}
\end{figure}

The Polaris Flare has a distance of about 352 pc  and  total mass of about 5500 M$_\odot$ \citep{2019ApJ...879..125Z}.  It is a high-latitude translucent cloud with little or no obvious star formation activity. It is expected to have the lowest level of background cloud emission  \citep{1990ApJ...353L..49H}. No OB stars nor YSOs were detected in this region. As shown in  Fig.\ref{fig:polaris}, the UV intensity G$_0$ of almost all region is smaller than 10. 

\subsection{UV Intensity $vs$ N(H$_2$) }
\label{subsec:uv_NH2}
\begin{figure}[h!]
\centering
\includegraphics[width=0.80\textwidth]{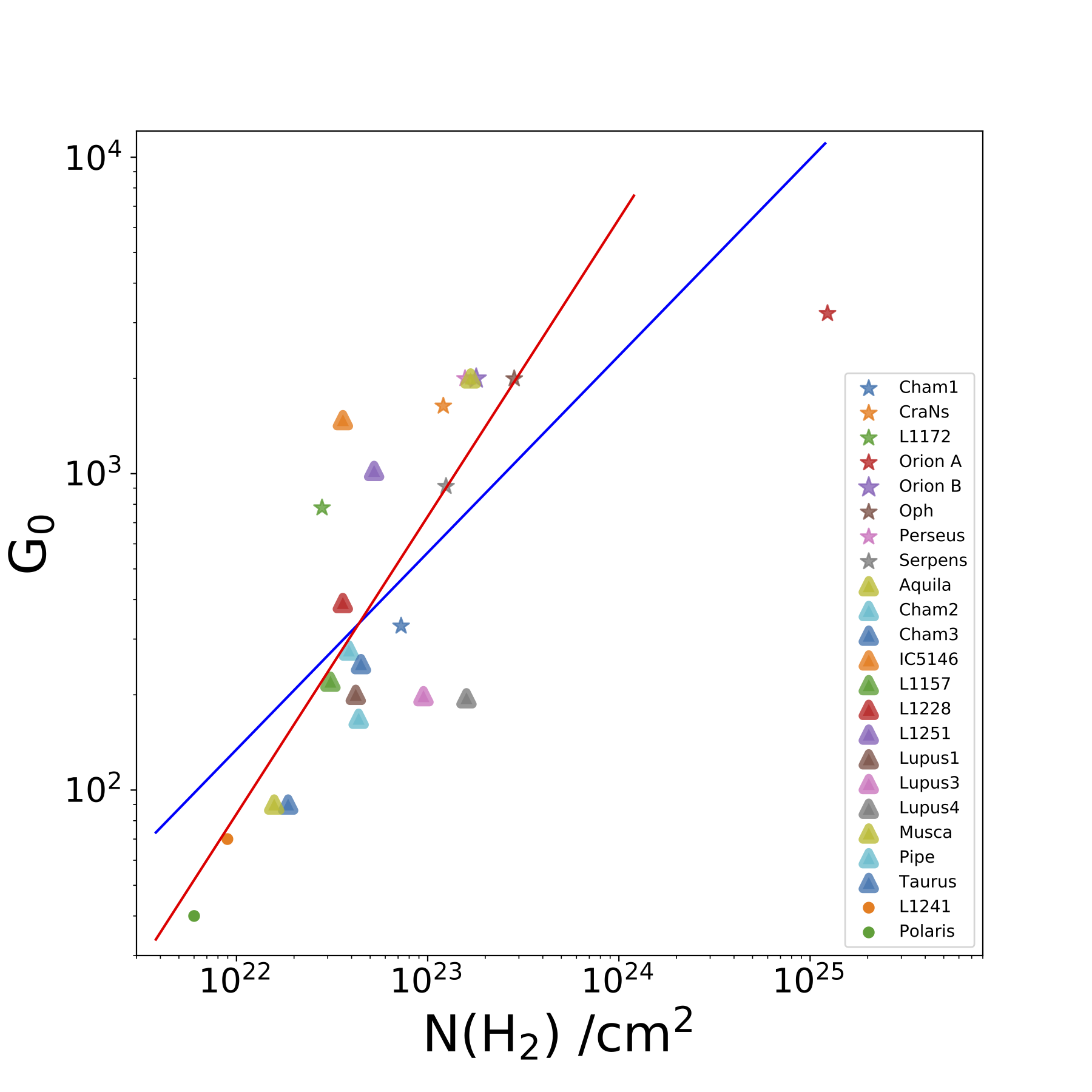}
\caption{The relationship between UV intensity $G_0$ and peak H$_2$ column density N(H$_2$). Filled star, triangle, and circle represent category I, II and III, respectively. The blue solid line represents linear fit of all clouds. The red solid line represents linear fit when Orion A is excluded. }
\label{fig:uv_Nh2}
\end{figure}

We presented a statistical result between UV intensity $G_0$ and peak H$_2$ column density N(H$_2$) of each molecular cloud  in Fig. \ref{fig:uv_Nh2}. The  correlation between $G_0$ and N(H$_2$) can be fitted with a linear function of  
\begin{equation}
    \rm log(G_0)=(0.62\pm0.12)log(N(H_2))-(11.56\pm2.87)  \>\>,\label{eq:logm1}
\end{equation}

for all complexes. It becomes 

\begin{equation}
    \rm log(G_0)=(0.94\pm0.17)log(N(H_2))-(18.76\pm3.91)  \>\>,\label{eq:logm2}
\end{equation}
when Orion A was not included.

\section{Discussion}
\label{sec:discussion}

In this paper, we obtained the distribution of UV radiation field  toward 23 regions of 14 molecular complexes in Gould Belt through dust radiative analysis.  

\subsection{Uncertainty Analysis}
The main assumption of such analysis is that the heating of dust was mainly due to UV radiation from massive stars. This has been demonstrated to be valid in the presence of massive stars, particularly, OB clusters \citep[e.g.,][]{1999ApJ...522..897L, 2003ApJ...587..262L}. In more quiescent regions, however, dynamic feedback through outflows and bubbles from low-mass stars have proven to be capable of sustaining the turbulence in, e.g.~Taurus \citep{2015ApJS..219...20L}, thus presumably may result in some dust heating.
However, even suppose the gas and dust are closely coupled, there is no sign of elevated temperatures close to low-mass YSOs in Taurus \citep{2008ApJ...680..428G}. We thus consider the heating of dust from low-mass YSOs to be minor.

In testing the model, we find that the incident angle affect more the absolute value of the derived UV field, rather than its distribution. 
To obtain robust results for the Gould belt sample in a systematic way, we set the incident angle to zero. Further investigation of the individual radiation geometry is warranted. 

\subsection {Compare with Star Formation Rate}
We obtain the total UV fluxes in Aquila, Cham I, Cham II, Cham III, IC5146, Lupus III, Lupus IV, Serpens,  Musca, $\rho$ Oph, Orion A, Orion B, Pipe, Perseus, and Taurus by adding up the DUSTY outputs for each region and compare them with the star formation rate in \citet{2010ApJ...724..687L,2014ApJ...782..114E}.

 UV emission is a direct tracer of the recent SFR since it traces the photospheric emission of young stars. The investigation of SFR of extragalaxies has been revolutionized with the observations of GALEX telescope \citep{2005ApJ...619L...1M}. By combining the UV data and Initial Mass Function (IMF), the relationship between SFR and UV luminosity \citep{2011ASPC..446...63H, 2012ARA&A..50..531K} can be described 
as 

\begin{equation}
    \rm log(\dot{M}_{\odot})=log(L_{\nu})-log(C) \>\>,\label{eq:logm}
\end{equation}
where $\dot{M_{\odot}}$ is SFR in unit of  10$^{-6}M_{\odot}$ year$^{-1}$, L$_{\nu}$(FUV) is FUV luminosity in unit of ergs s$^{-1}$  , log(C)= 43.35 is conversion constant. 

\begin{figure}[ht!]
\centering
 \includegraphics[width=0.6\textwidth,height=0.6\textwidth,trim={{0.05\textwidth} {0.25\textwidth}  {0.05\textwidth}  {0.1\textwidth} },clip]{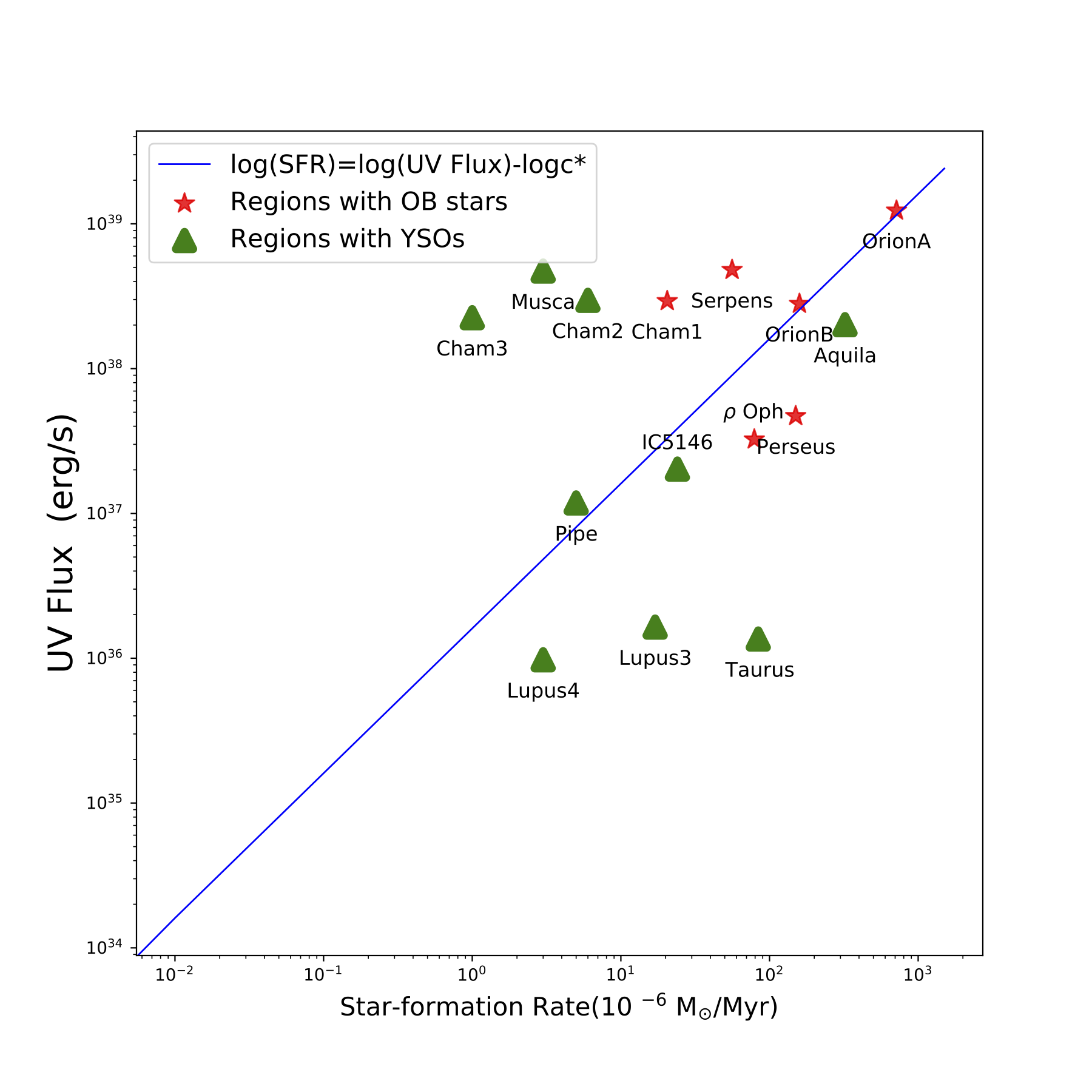}
 \caption{The relationship between total UV flux  and star formation rate in 15 molecular clouds including Lupus, Corona Australis, Pipe, Musca, $\rho$ Oph, Perseus, Aquila, Orion A, Orion B, Serpens, IC 5146, Chamaleon. The blue line indicates the formula of equation \ref{eq:logm}. Regions with massive OB stars fit the blue line better than regions with YSOs.
 }
 \label{fig:sr-intensity}
 \end{figure}

The relationship between UV intensity and SFR for 15 molecular regions of Gould Belt is presented in Fig. \ref{fig:sr-intensity}. The blue line denotes the correlation between UV flux by OB stars and star formation rate. The correlation conforms to the general expectation.
Regions with prominent OB clusters  tend to be more consistent with the expectation.
The scatter is bigger where there is no OB star. Different regions in the same molecular cloud tend to have similar UV fields, but different star formation rate.
The less massive YSOs seem to have little effect on the UV distribution.

\section{Summary}
\label{sec:summary}
By interpreting dust continuum data through radiative transfer analysis, we obtained UV intensity distribution toward 23 molecular regions in the Gould Belt. The main results of this study are summarized as follows, 

\begin{enumerate}
\item  The UV intensity G$_0$ of molecular clouds ranges from 1 to  over 1000, relative to the Habing interstellar field. 

\item The UV distribution in the majority of the molecular regions shows a tight correlation with that of OB stars and/or YSOs.

\item The  UV intensity of 10 molecular  regions conforms to an expected linear correlation with the SFR. 

\end{enumerate}

\normalem
\begin{acknowledgements}
This work is supported by the National Natural Science Foundation of China (NSFC) grant No. 11988101, No. 11725313, No. 11721303, No. U1731238, the International Partnership Program of Chinese Academy of Sciences grant No.114A11KYSB20160008, the National Key R\&D Program of China No. 2016YFA0400702, the Guizhou Provincial Science and Technology Foundation (Nos. [2016]4008, [2017]5726-37,), the Foundation of Guizhou Provincial Education Department (No. KY (2020) 003). This research was carried out in part at the Jet Propulsion Laboratory, California Institute of Technology, under contract with the National Aeronautics and Space Administration. This research made use of {\sc APLpy}, an open-source plotting package for Python\citep{aplpy2012,aplpy2019}.

\end{acknowledgements}


\appendix
\section{UV Distribution of the Gould Belt Complexes}

\subsection{Cepheus}
Located at a high declination, the Cepheus molecular complex includes many regions that have loose association with compact dark clouds. HGBS contains 5 regions in Cepheus molecular complex, Cep 1157 (also known as L1157, the same below), Cep 1172, Cep 1228, Cep 1241 and Cep 1251. The distance of the five regions are considered to be 200-300 pc. The masses of L1157, L1172, L1228, L1241, L1251 are 1400 M$_{\odot}$, 1900 M$_{\odot}$, 1600 M$_{\odot}$, 3200 M$_{\odot}$ and 1800 M$_{\odot} $ \citep{2020ApJ...904..172D}. L1172 is the host of the bright NGC 7023 reflection nebula, which contains the bright B star, HD200775 in these five regions. L1241 and L1251 lie within the Cepheus Flare Shell. L1241 is the only one region without YSOs nor OB star. Some YSOs are found in L1251.  The L1228 locates at the edge of the Cepheus Flare Shell, while L1157 and L1172 locate outside to the Cepheus Flare Shell. 

\begin{figure}[h!]
 \centering
 \includegraphics[width=0.80\textwidth,height=0.80\textwidth,trim={{0.20\textwidth} {0.60\textwidth}  {0.05\textwidth}  {0.05\textwidth} },clip]{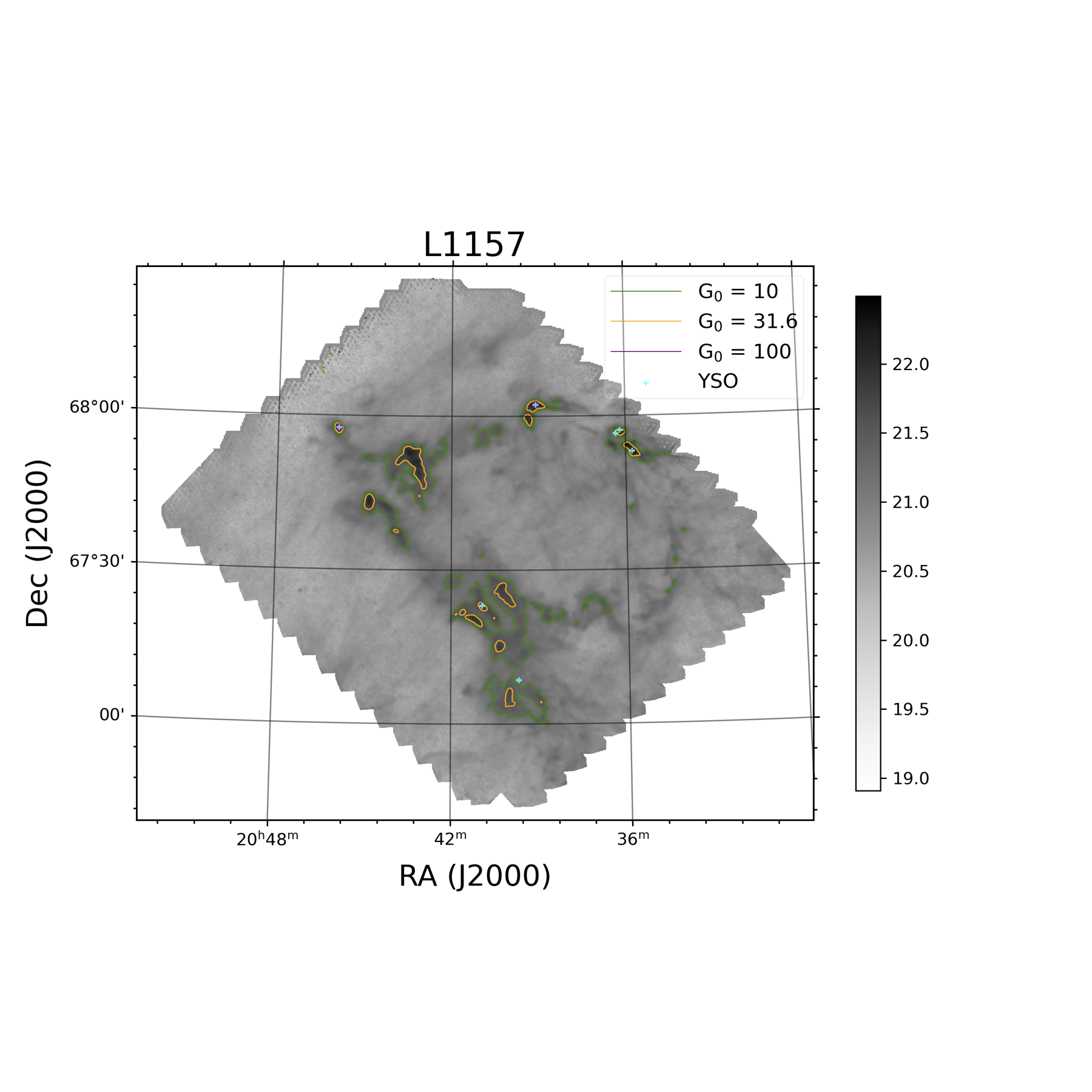}
 \caption{Contours of UV intensity overlaid on H$_2$ column density map of the L1157 region. The labels of the figure are the same as that in Fig. \ref{fig:aquila}.}
 \label{fig:l1157}
 \end{figure}

\begin{figure}[h!]
 \centering
 \includegraphics[width=1.10\textwidth,height=0.80\textwidth,trim={{0.10\textwidth} {0.4\textwidth}  {0.05\textwidth}  {0.05\textwidth} },clip]{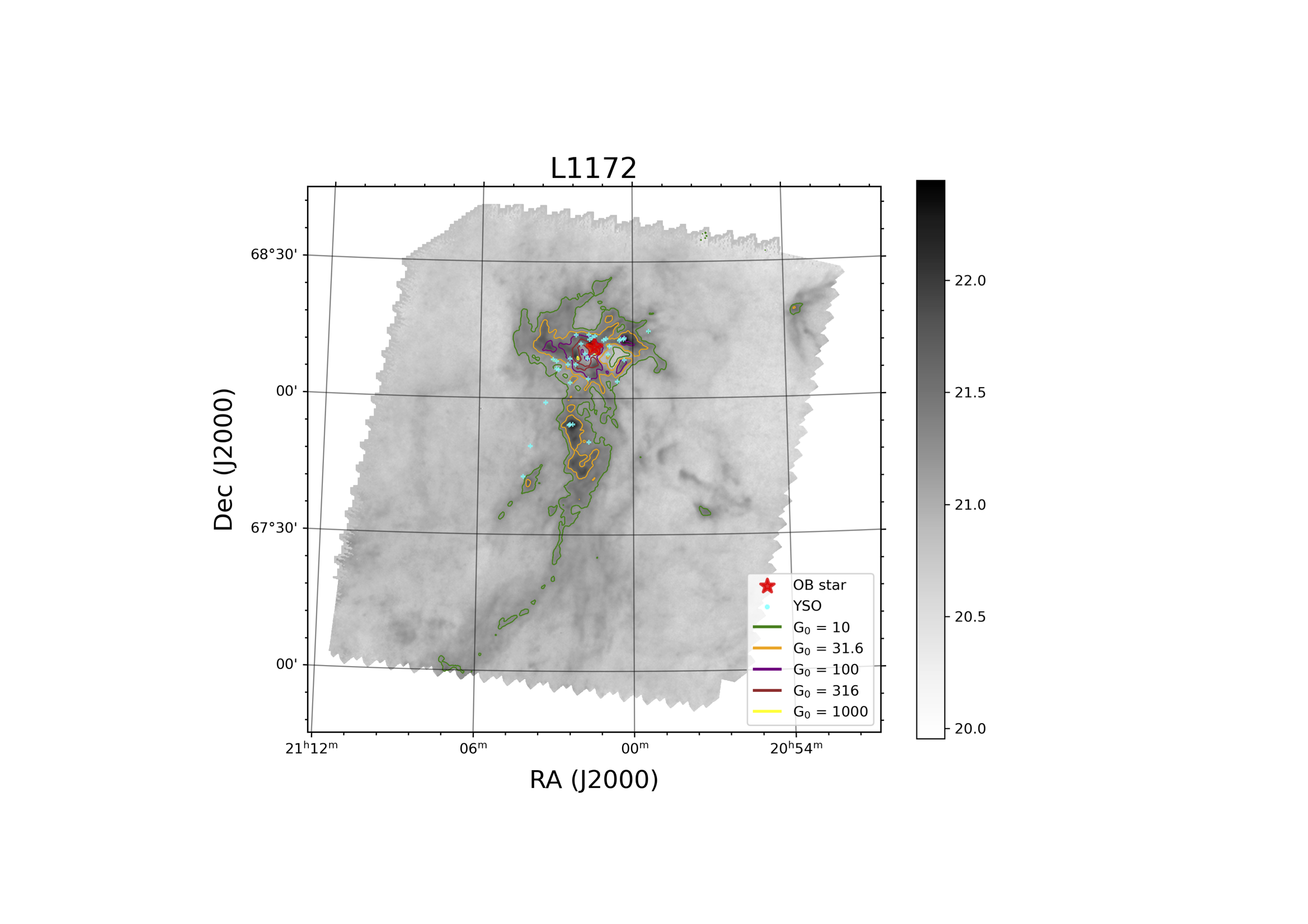}
 \caption{Contours of UV intensity overlaid on H$_2$ column density map of the L1172 region. The labels of the figure are the same as that in Fig. \ref{fig:aquila}. The red star implies the location of OB star in this region. }
 \label{fig:l1172}
 \end{figure}

\begin{figure}[h!]
 \centering
 \includegraphics[width=1.10\textwidth,height=0.60\textwidth,trim={{0.10\textwidth} {0.65\textwidth}  {0.05\textwidth}  {0.1\textwidth} },clip]{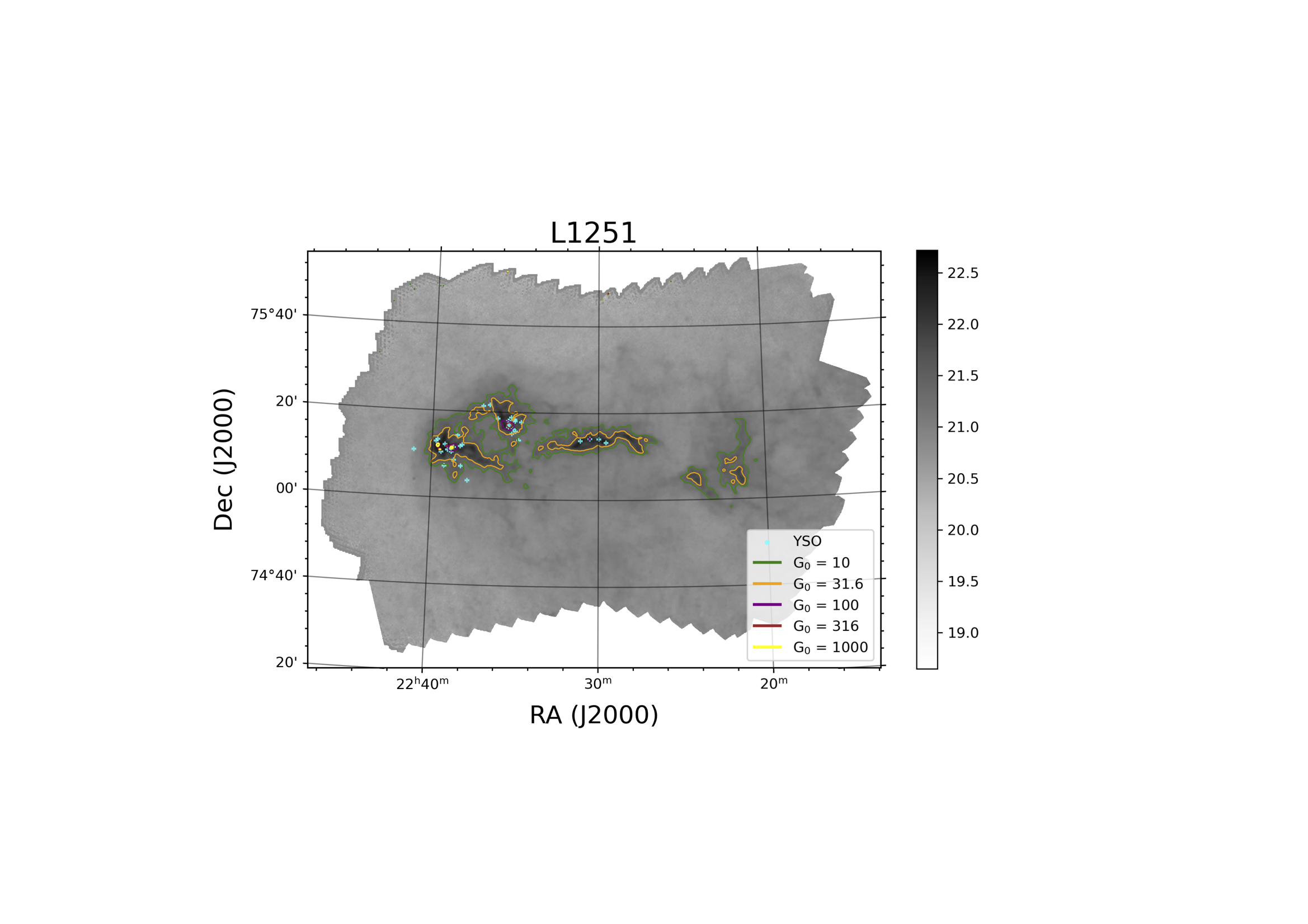}
 \caption{Contours of UV intensity overlaid on H$_2$ column density map of the L1251 region. The labels of the figure are the same as that in Fig. \ref{fig:aquila}.}
 \label{fig:l1251}
 \end{figure}

The radiation field distribution of these five regions are quite different depending on the existence of massive stars and YSOs. As shown in Fig.\ref{fig:l1172}, the UV intensity of L1172 around HD 200775 can exceed 1000 G$_0$. For the L1157 ( Fig. \ref{fig:l1157}), L1228 (Fig.\ref{fig:l1172}) and L1251 (Fig.\ref{fig:l1251}) regions, obvious higher UV intensity can be found around the YSOs. The UV intensity in L1241 is smaller than 31.6 G$_0$ while there are no massive stars nor YSOs in this region.

 \begin{figure}[h!]
 \centering
 \includegraphics[width=1.10\textwidth,height=0.80\textwidth,trim={{0.10\textwidth} {0.4\textwidth}  {0.05\textwidth}  {0.05\textwidth} },clip]{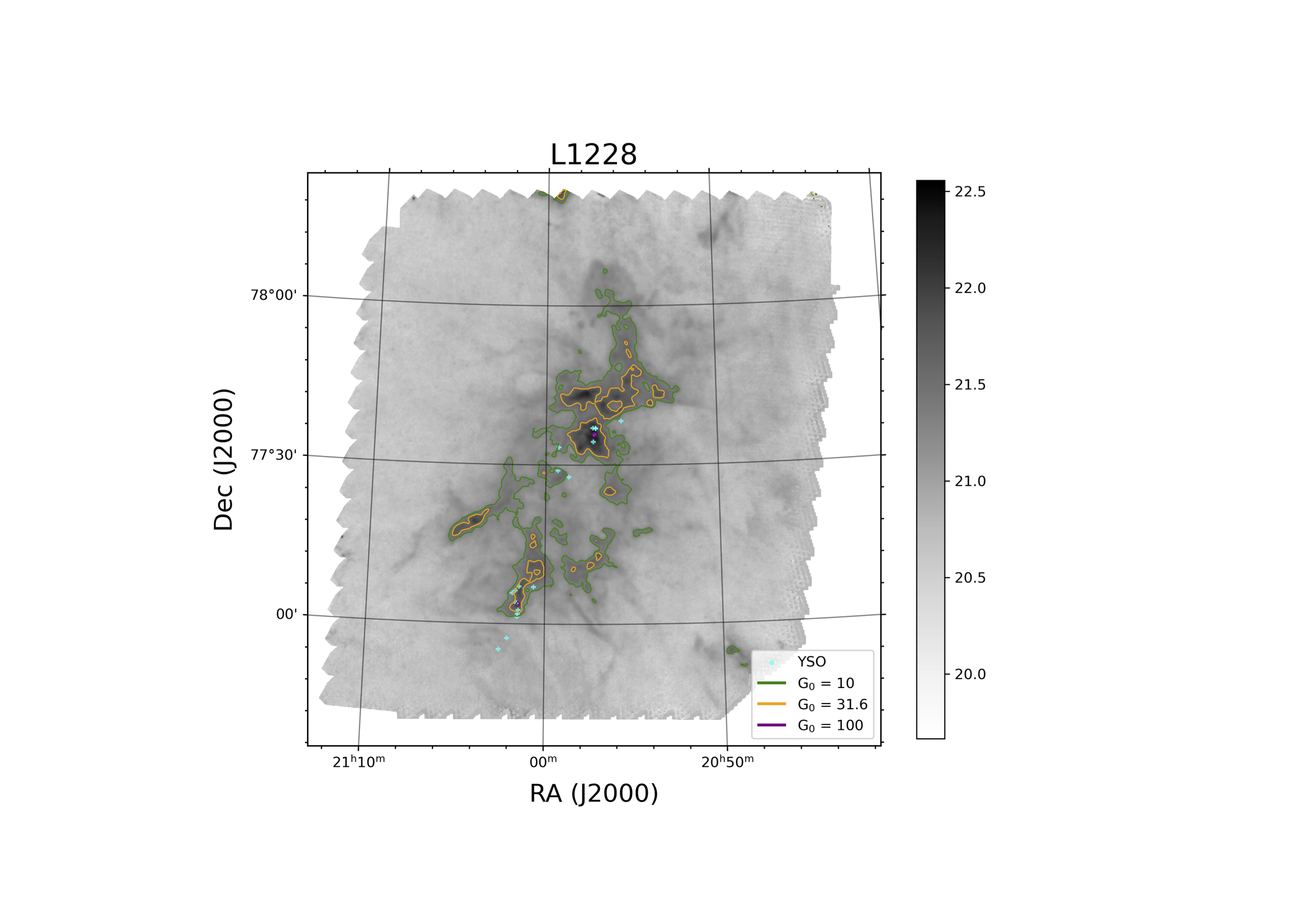}
 \caption{Contours of UV intensity overlaid on H$_2$ column density map of the L1228 region. The labels of the figure are the same as that in Fig. \ref{fig:aquila}.}
 \label{fig:l1228}
 \end{figure}

 \begin{figure}[h!]
 \centering
 \includegraphics[width=1.10\textwidth,height=0.80\textwidth,trim={{0.10\textwidth} {0.10\textwidth}  {0.05\textwidth}  {0.05\textwidth} },clip]{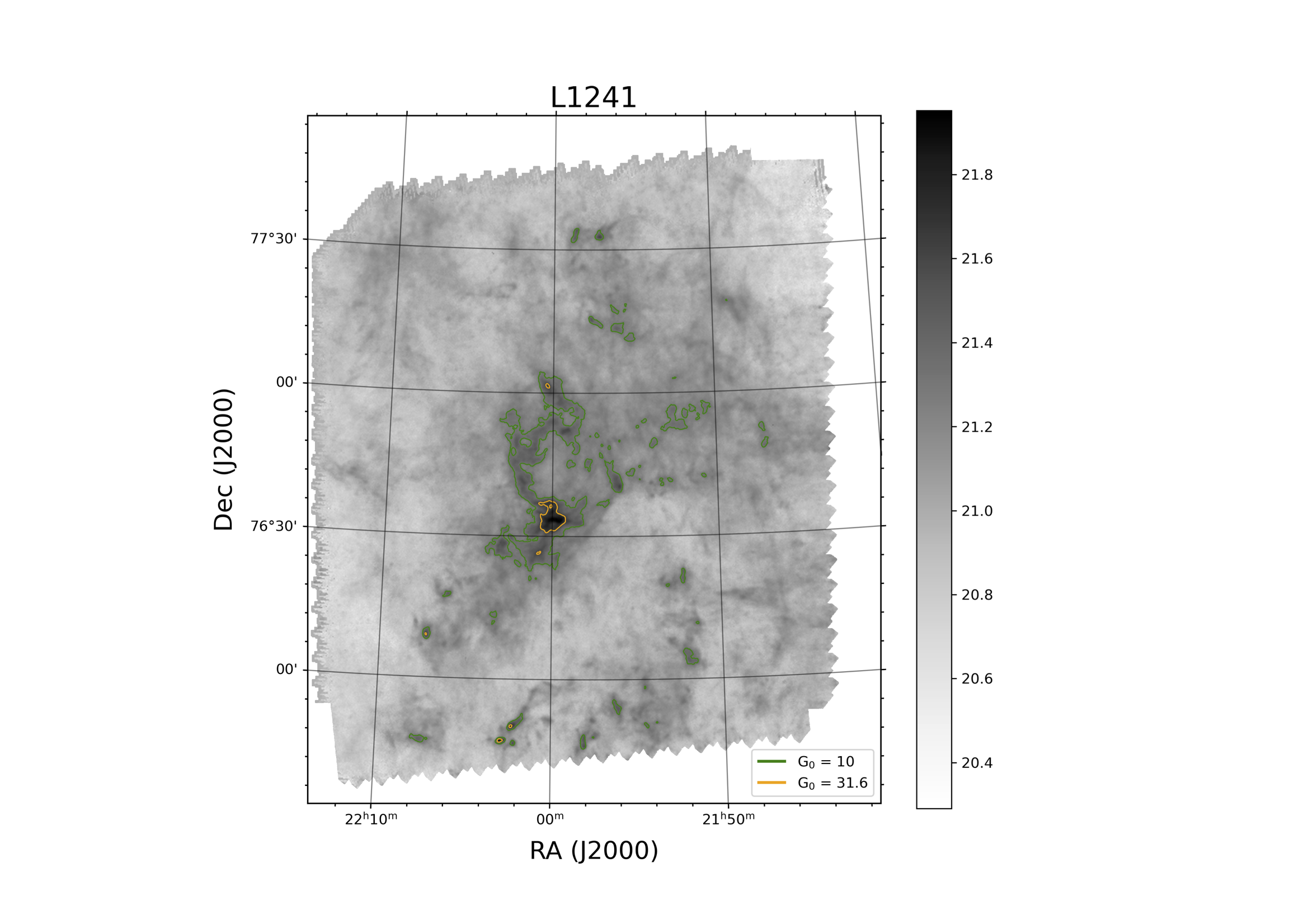}
 \caption{ Contours of UV intensity overlaid on H$_2$ column density map of the L1241 region. The labels of the figure are the same as that in Fig. \ref{fig:aquila}. There is no YSOs nor OB star in this region.}
 \label{fig:l1241}
 \end{figure}

\subsection{Chamaleon}
Chamaleon is a nearby low-mass star-forming region containing Cham I, Cham II and Cham III. The distance from the regions in Chamaleon is about 150 pc \citep{2014ApJ...782..114E}. The entire Cham I region covers an area about ~5 deg{$^2$}, containing about 200 known low-mass YSOs, making it one of the closest and richest star-forming regions. The total mass of this cloud is about 482 M$_{\odot}$, one third of which are dense gas \citep{2014ApJ...782..114E}. Though there is abundant low-mass YSOs in Cham I, there is only one B star: HD 97300 in the northern part of the cloud \citep{2012A&A...545A.145W}. 

Cham II region contains a smaller number($\sim$60) of YSOs compared to the Cha I region. The size of Cham II is about 1.78\,pc. The total mass of Cham II is about 637 M$_{\odot}$ while one tens is dense gas. As the largest cloud  of the three regions with a total mass of 746 M$_{\odot}$, Cham III contain little dense gas and only a few YSOs\citep{2014A&A...568A..98A}. 

\begin{figure}[h!]
 \centering
 \includegraphics[width=0.90\textwidth,height=0.70\textwidth,trim={{0.10\textwidth} {0.15\textwidth}  {0.05\textwidth}  {0.1\textwidth} },clip]{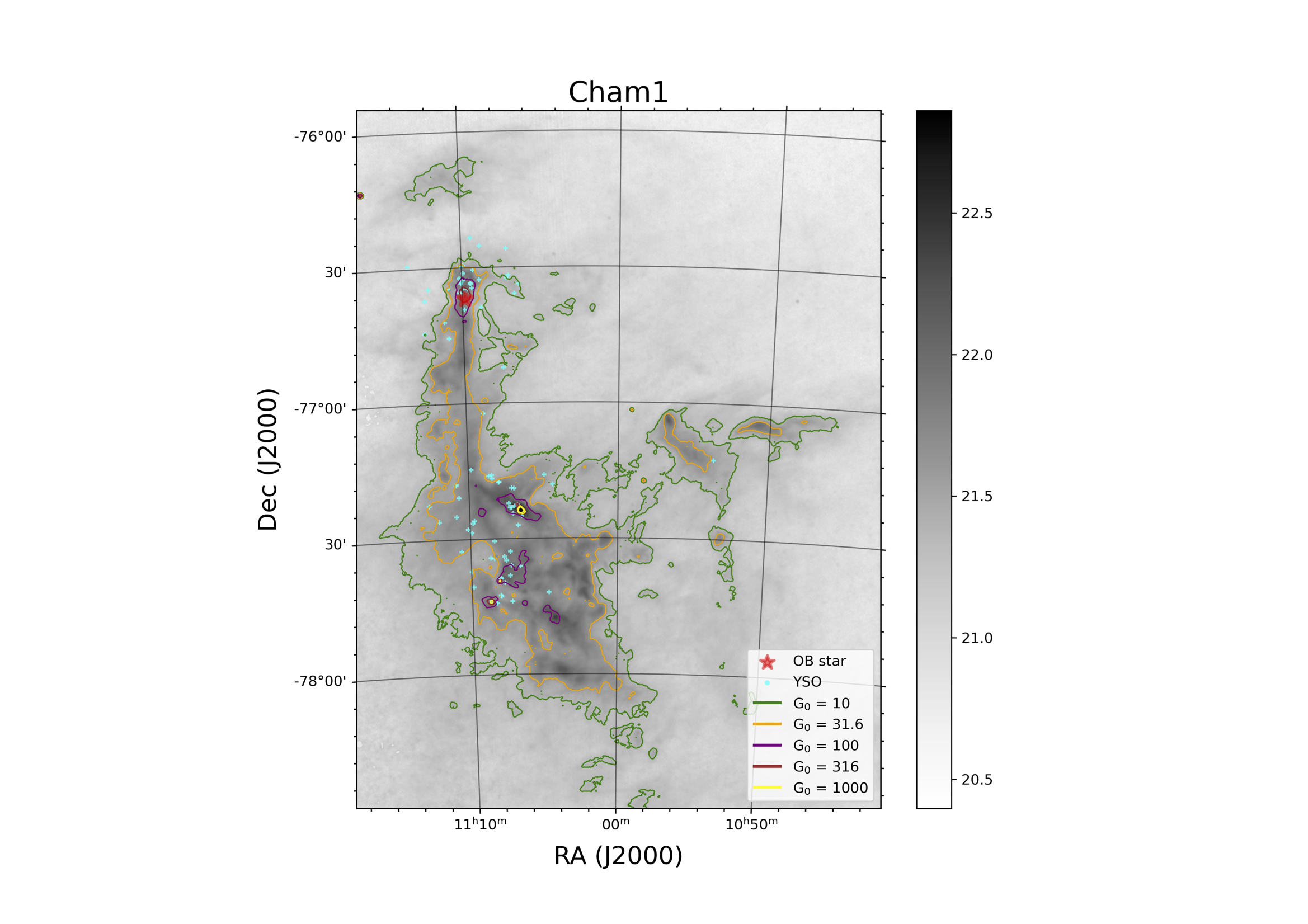}
 \caption{Contours of UV intensity overlaid on H$_2$ column density map of the Cham I region. The labels of the figure are the same as that in Fig. \ref{fig:l1172}.}
 \label{fig:cham1}
 \end{figure}

\begin{figure}[h!]
 \centering
 \includegraphics[width=1.10\textwidth,height=0.80\textwidth,trim={{0.01\textwidth} {0.35\textwidth}  {0.05\textwidth}  {0.1\textwidth} },clip]{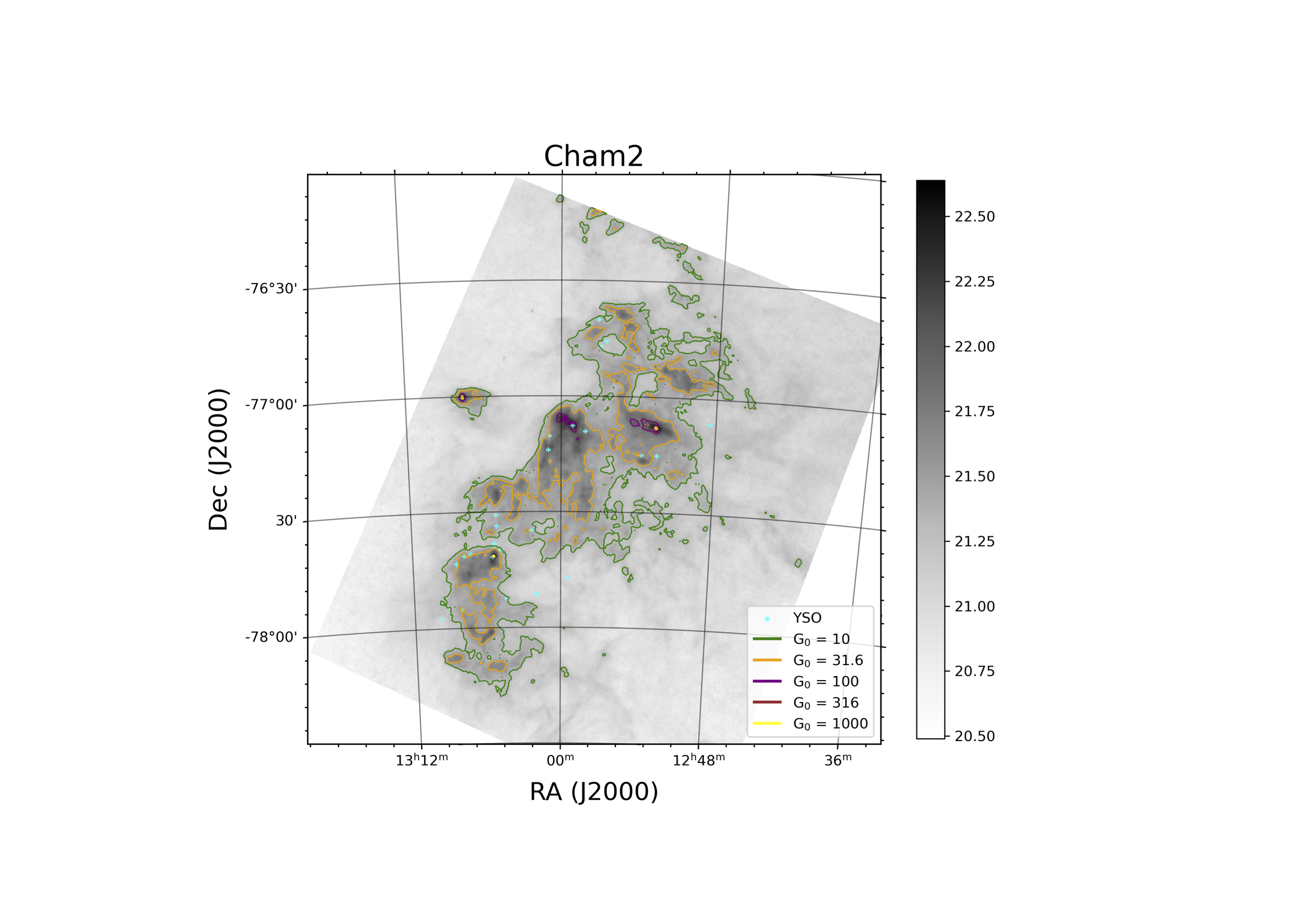}
 \caption{Contours of UV intensity overlaid on H$_2$ column density map of the Cham II region. The labels of the figure are the same as that in Fig. \ref{fig:aquila}.}
 \label{fig:cham2}
 \end{figure}

\begin{figure}[h!]
 \centering
 \includegraphics[width=1.10\textwidth,height=0.80\textwidth,trim={{0.01\textwidth} {0.35\textwidth}  {0.05\textwidth}  {0.1\textwidth} },clip]{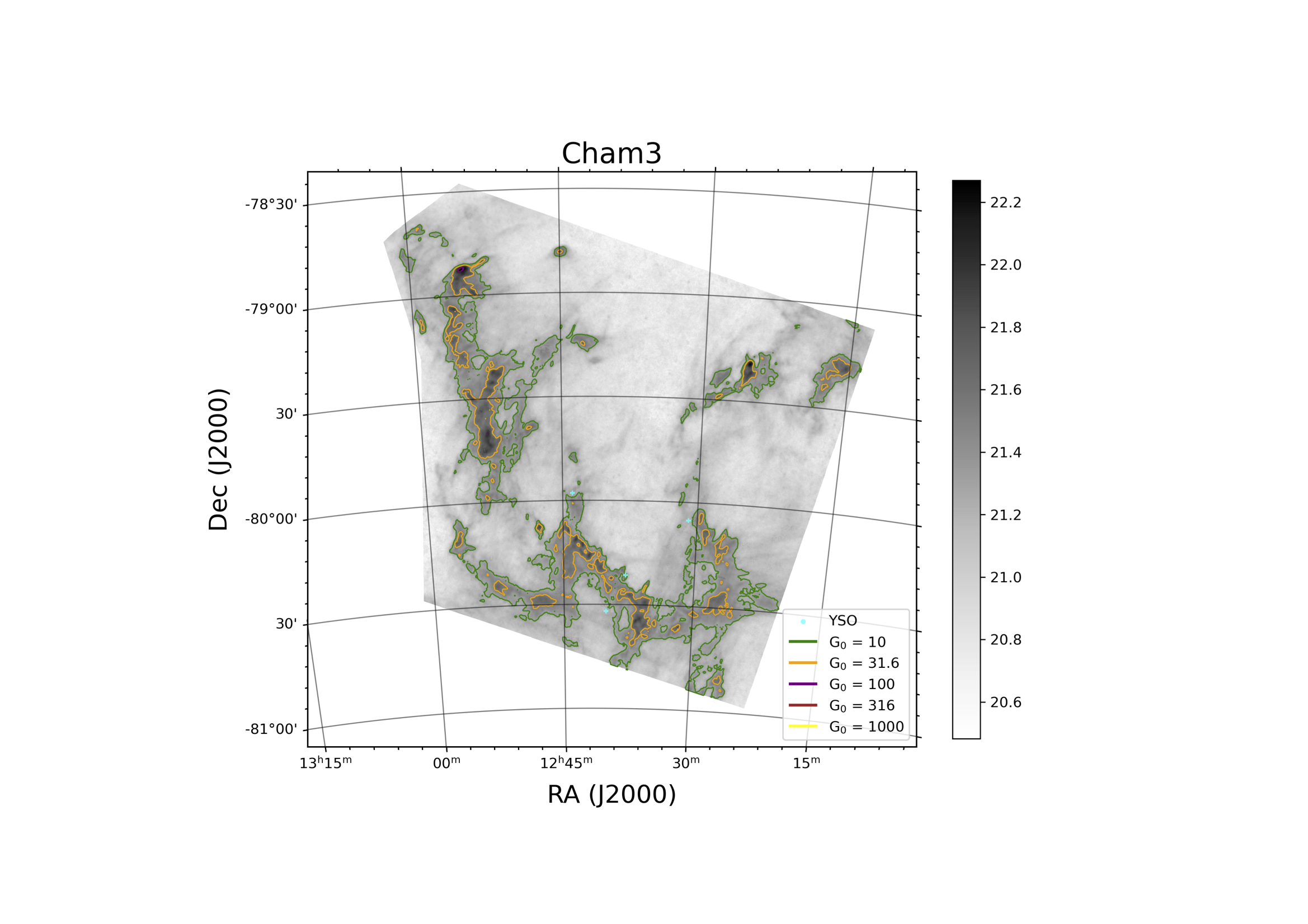}
 \caption{Contours of UV intensity overlaid on H$_2$ column density map of the Cham III region. The labels of the figure are the same as that in Fig. \ref{fig:aquila}.}
 \label{fig:cham3}
 \end{figure}

Due to the differences in containing YSOs and dense gas, the UV intensity distribution of Cham I, Cham II and Cham III are expected to vary significantly. As shown in Fig. \ref{fig:cham1}, \ref{fig:cham2}, and \ref{fig:cham3}, the maximum UV intensity value decrease from Cham I to Cham III, which is proportional with the existence of OB stars. 

\subsection{CraNS}
\begin{figure}[h!]
 \centering
 \includegraphics[width=0.80\textwidth,height=0.70\textwidth,trim={{0.30\textwidth} {0.80\textwidth}  {0.05\textwidth}  {0.1\textwidth} },clip]{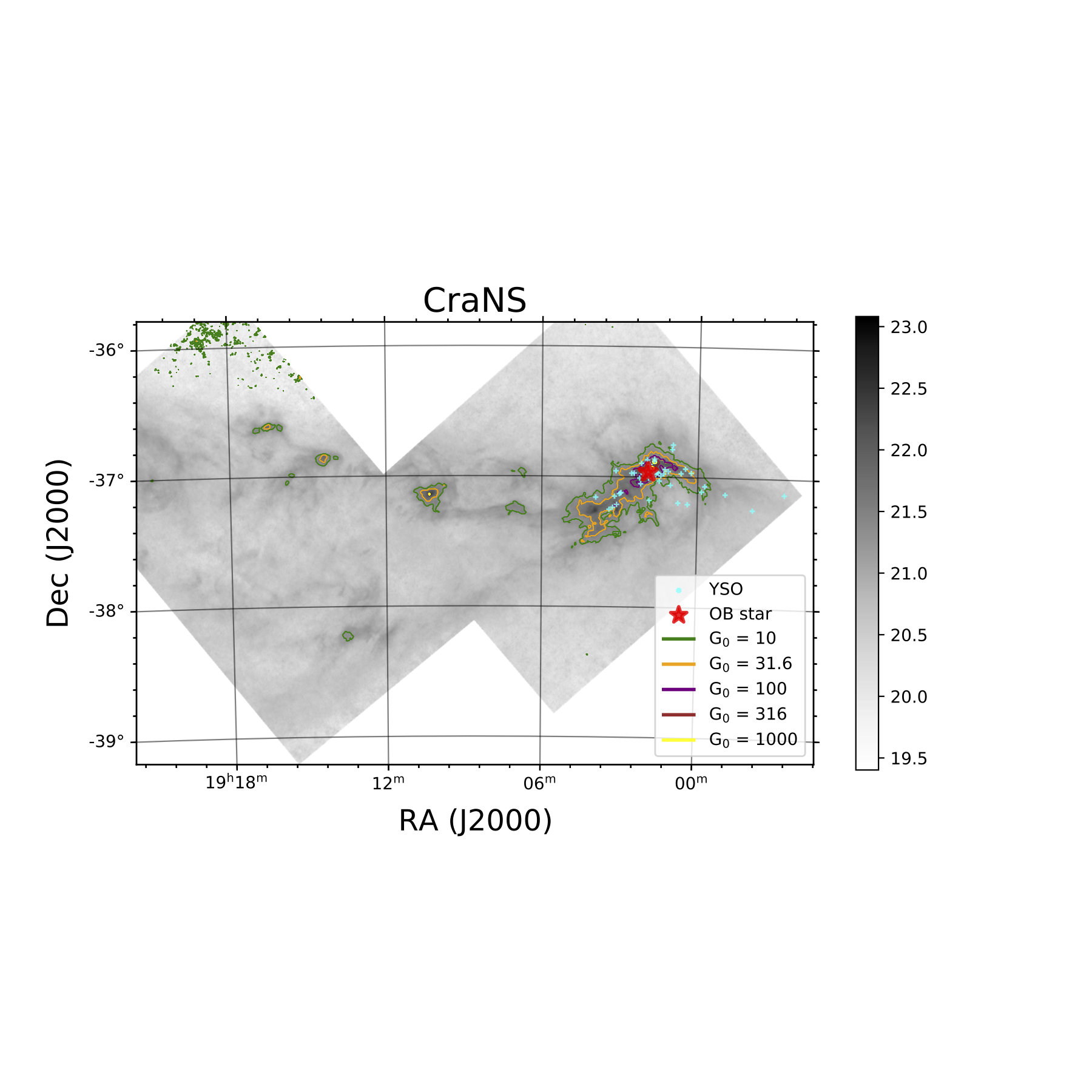}
 \caption{Contours of UV intensity overlaid on H$_2$ column density map of the CraNS region. The labels of the figure are the same as that in Fig. \ref{fig:aquila}.}
 \label{fig:crans}
 \end{figure}

With distance of around 130\,pc and being out of the Galactic plane, the CraNS (Corona Australias) molecular cloud is a low-mass star-forming region. The total mass of this region is about 279 M$_{\odot}$, half of which is dense gas \citep{2014ApJ...782..114E}. There is an OB star(V$^*$ R CrA) and a cluster of YSOs in this region \citep{2018A&A...615A.125B}. As shown in Fig. \ref{fig:crans}, the derived UV intensity G$_0$ around the OB star can exceed 1000.

\subsection{IC 5146}
\begin{figure}[h!]
 \centering
 \includegraphics[width=0.80\textwidth,height=0.70\textwidth,trim={{0.10\textwidth} {0.3\textwidth}  {0.05\textwidth}  {0.1\textwidth} },clip]{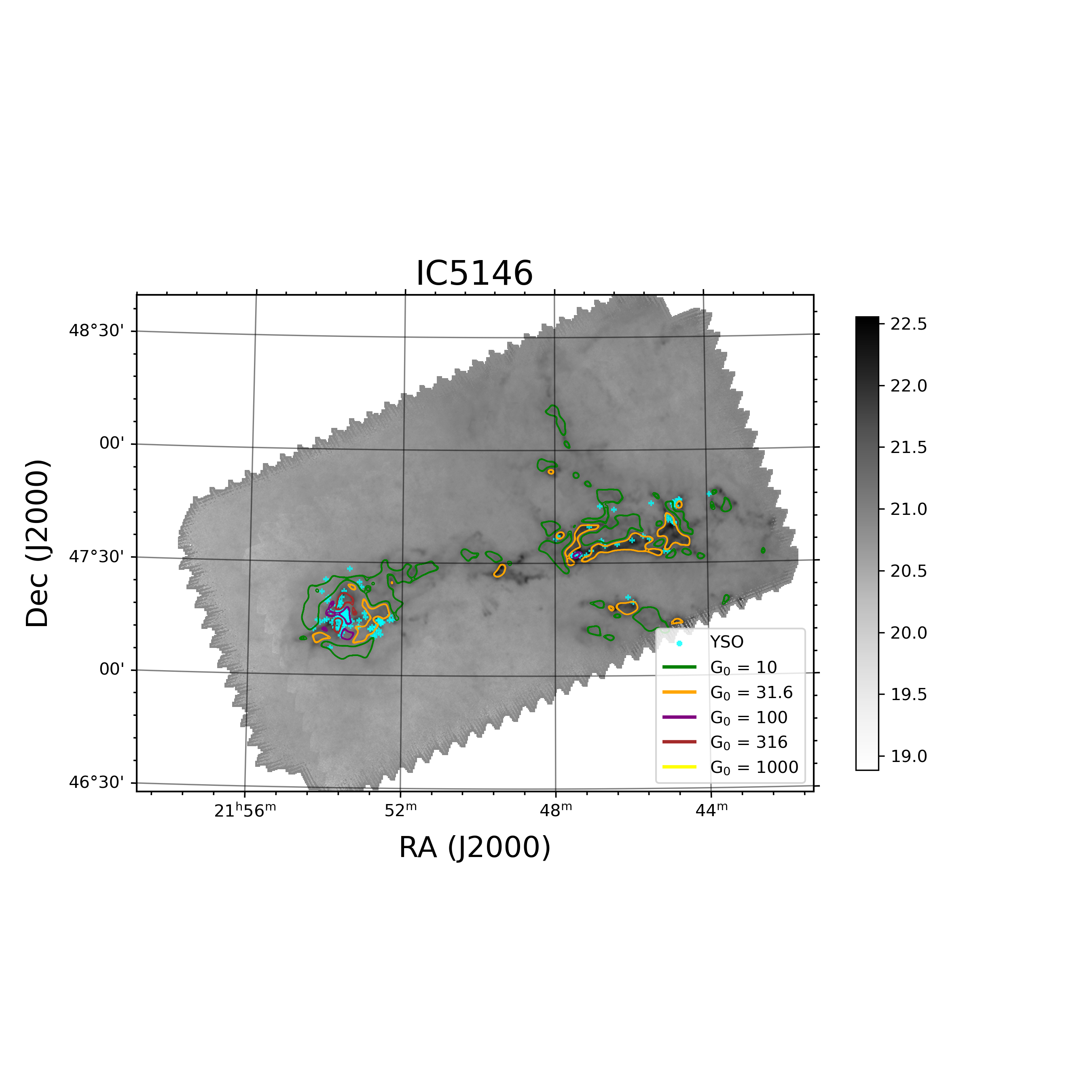}
 \caption{Contours of UV intensity overlaid on H$_2$ column density map of the IC 5146 region. The labels of the figure are the same as that in Fig. \ref{fig:aquila}.}
 \label{fig:ic5146}
 \end{figure}

With a distance of $\sim$ 460 pc, IC 5146 region covers an area of $\sim$ 3.1 $\times$ 2.5 deg{$^2$}. The total mass of this cloud is about 3.7 $\times$ 10 $^3$ M{$_ \odot $} \citep{2019A&A...621A..42A}. There is little dense gas nor OB star in this region \citep{2014ApJ...782..114E}. Abundant YSOs exist in this region. We present UV intensity distribution for IC 5146 in Fig. \ref{fig:ic5146}. The UV intensity G$_0$ can reach 31.6 around the YSOs.  

\subsection{Lupus}
The distance of Lupus molecular complex is about 189 pc \citep{2019ApJ...879..125Z}. The HGBS surveyed three clouds in this complex: Lupus I, Lupus III and Lupus IV cloud. 
Among the three clouds, the Lupus I cloud is the youngest one. The mass of Lupus I is about 512 M$_{\odot}$. Lup III is the most evolved cloud with mass of $\sim$ 912 M$_{\odot}$. Lupus IV cloud has a middle property between Lupus I and III cloud. The mass of dense gas in Lupus IV cloud is 50 M$_{\odot}$, accounting for about one quarter of the cloud mass \citep{2014ApJ...782..114E}. These three clouds contain YSOs but there are no massive stars inside. 

\begin{figure}[h!]
 \centering
 \includegraphics[width=0.80\textwidth,height=0.80\textwidth,trim={{0.01\textwidth} {0.2\textwidth}  {0.05\textwidth}  {0.1\textwidth} },clip]{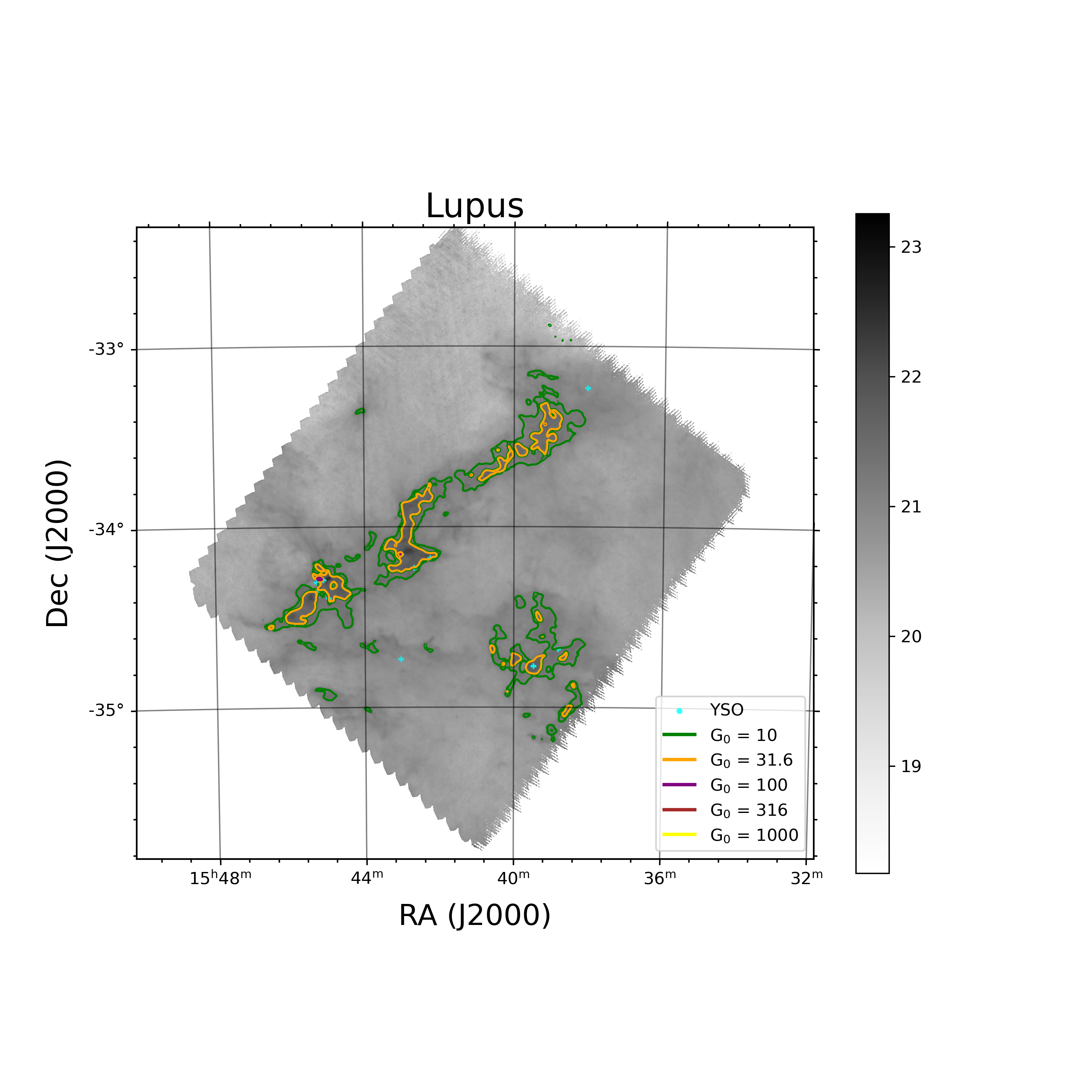} 
 \caption{Contours of UV intensity overlaid on H$_2$ column density map of the Lupus I region. The labels of the figure are the same as that in Fig. \ref{fig:aquila}.} 
 \label{fig:lupus1}
 \end{figure}

\begin{figure}[h!]
 \centering
 \includegraphics[width=0.80\textwidth,height=0.80\textwidth,trim={{0.01\textwidth} {0.22\textwidth}  {0.05\textwidth}  {0.1\textwidth} },clip]{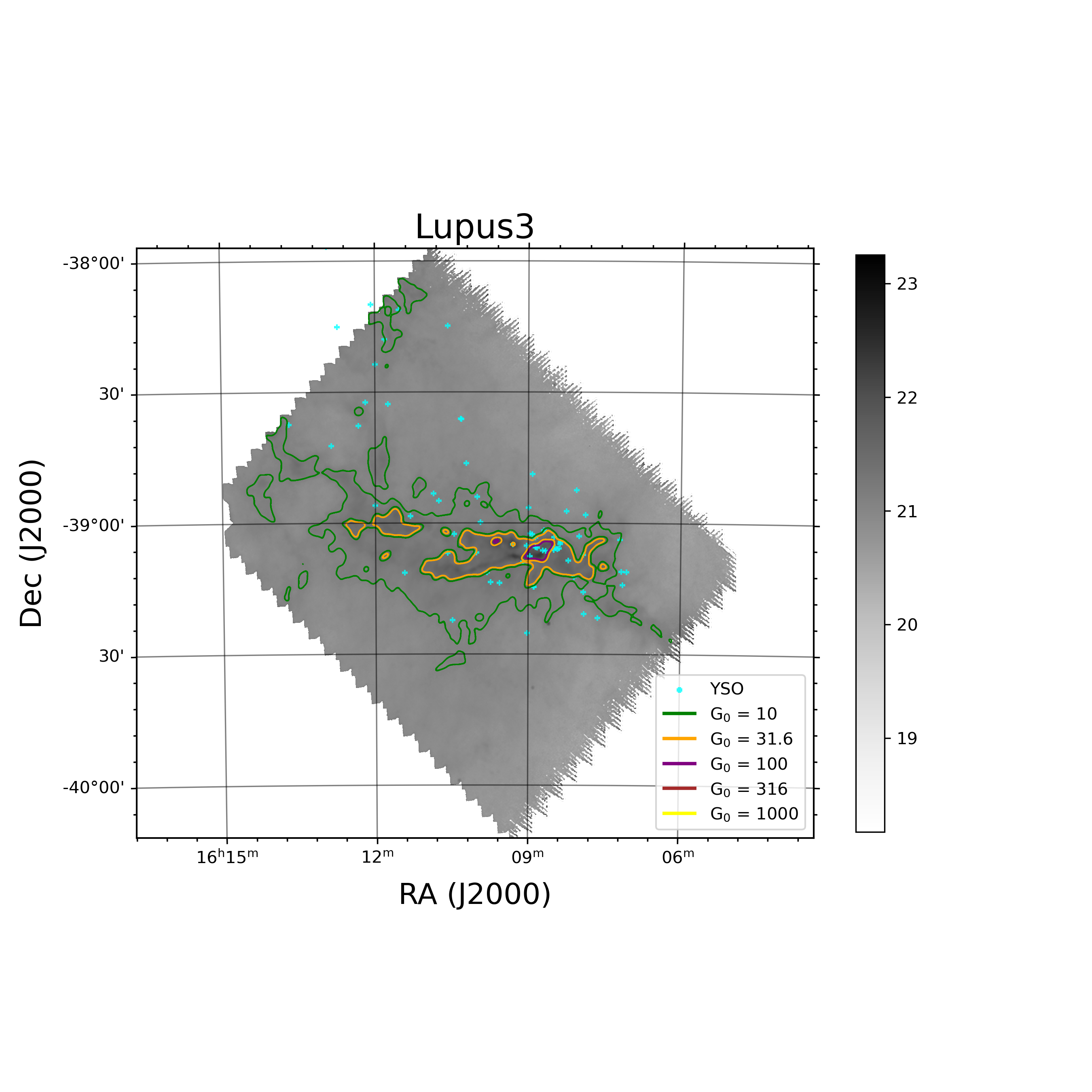}
 \caption{Contours of UV intensity overlaid on H$_2$ column density map of the Lupus III region. The labels of the figure are the same as that in Fig. \ref{fig:aquila}.}
 \label{fig:lupus3}
 \end{figure}

\begin{figure}[h!]
 \centering
 \includegraphics[width=0.80\textwidth,height=0.80\textwidth,trim={{0.01\textwidth} {0.3\textwidth}  {0.05\textwidth}  {0.1\textwidth} },clip]{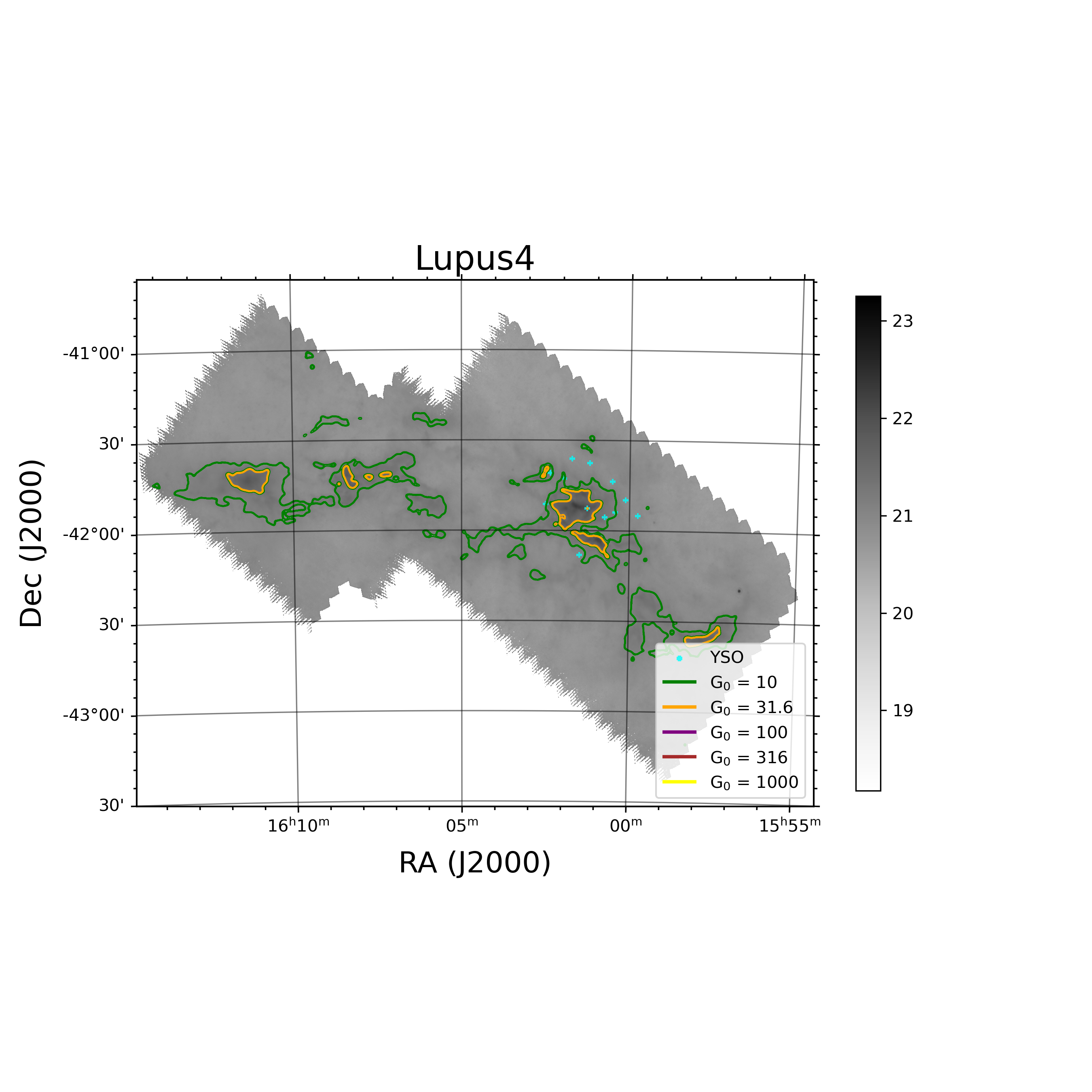}
 \caption{Contours of UV intensity overlaid on H$_2$ column density map of the Lupus IV region. The labels of the figure are the same as that in Fig. \ref{fig:aquila}.}
 \label{fig:lupus4}
 \end{figure}

As shown in Fig \ref{fig:lupus1}, \ref{fig:lupus3} and  \ref{fig:lupus4}, the UV intensity distribution correlate with the existences of YSOs. 

\subsection{Musca}
\begin{figure}[h!]
 \centering
 \includegraphics[width=0.80\textwidth,height=0.80\textwidth,trim={{0.01\textwidth} {0.1\textwidth}  {0.05\textwidth}  {0.1\textwidth} },clip]{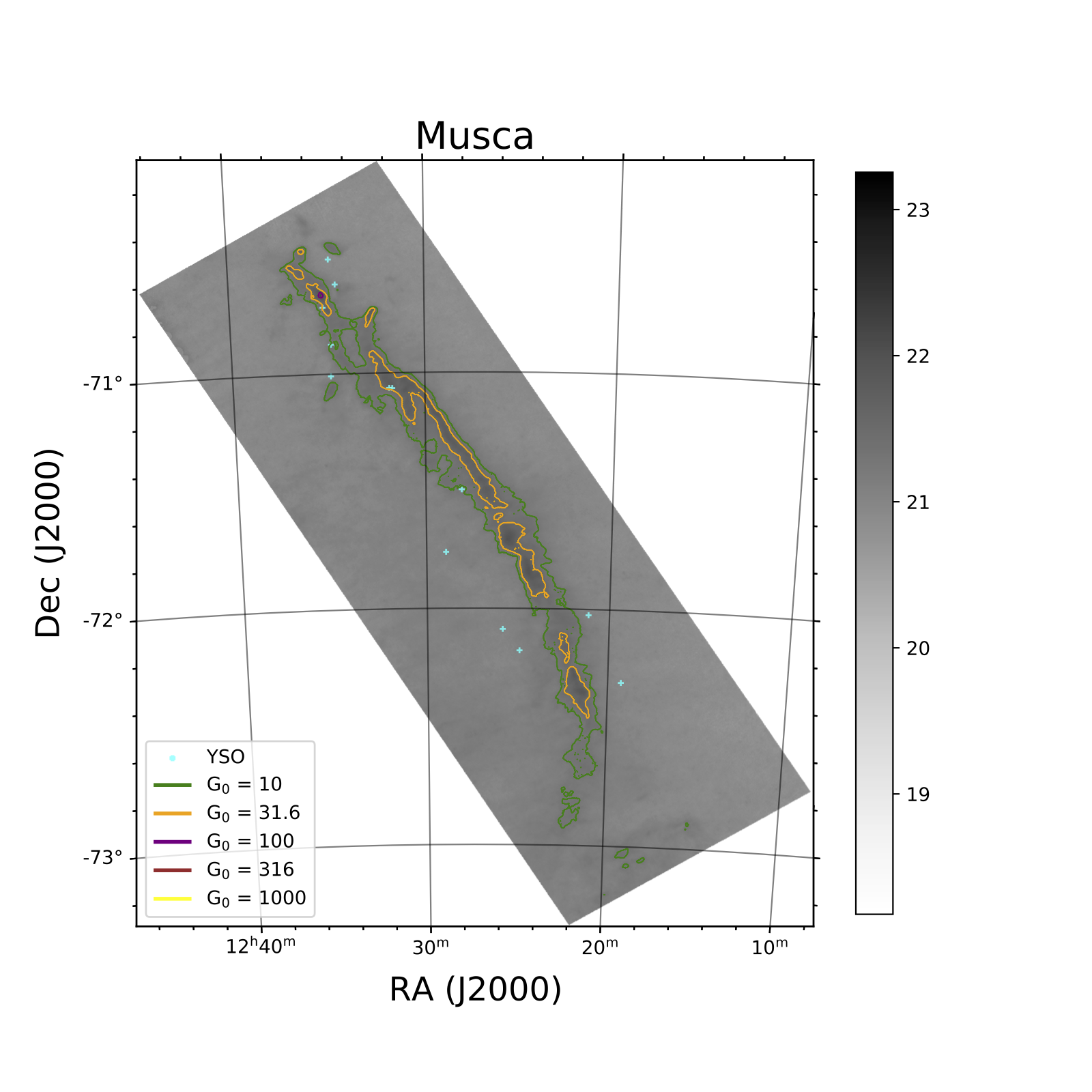}
 \caption{Contours of UV intensity overlaid on H$_2$ column density map of the Musca region. The labels of the figure are the same as that in Fig. \ref{fig:aquila}.}
 \label{fig:musca}
 \end{figure}

With distance of $\sim$ 200 pc, the Musca cloud is a 10.5 pc long filament with low-mass star formation \citep{2016A&A...590A.110C}. The mass of Musca molecular cloud is about 335 M$_{\odot}$ \citep{2014ApJ...782..114E}. There are no massive stars in this region. It is clear from Fig. \ref{fig:musca}, UV intensity increases in some dense regions.

\subsection{$\rho$ Oph}
With distance of 125\,pc \citep{2014ApJ...782..114E}, the $\rho$ Oph molecular cloud is one of the most conspicuous nearby regions where low and intermediate-mass star formation is taking place \citep{1992lmsf.book..159W}. The total mass of the $\rho$ Oph cloud is about 3128 M$_{\odot}$, one third of which is dense gas. The $\rho$ Oph cloud consists of two massive, centrally condensed cores, L1688 and L1689 \citep{1989ApJ...338..902L}. Being different from L1689 with little star formation activity, L1688 harbors a rich cluster of YSOs at various evolutionary stages and is distinguished by high star-formation efficiency \citep{1983ApJ...274..698W}. Two OB stars(HD 147889 and $\rho$ Oph A) are found in this region. 

\begin{figure}[h!]
 \centering
 \includegraphics[width=0.80\textwidth,height=0.80\textwidth,trim={{0.01\textwidth} {0.30\textwidth}  {0.05\textwidth}  {0.1\textwidth} },clip]{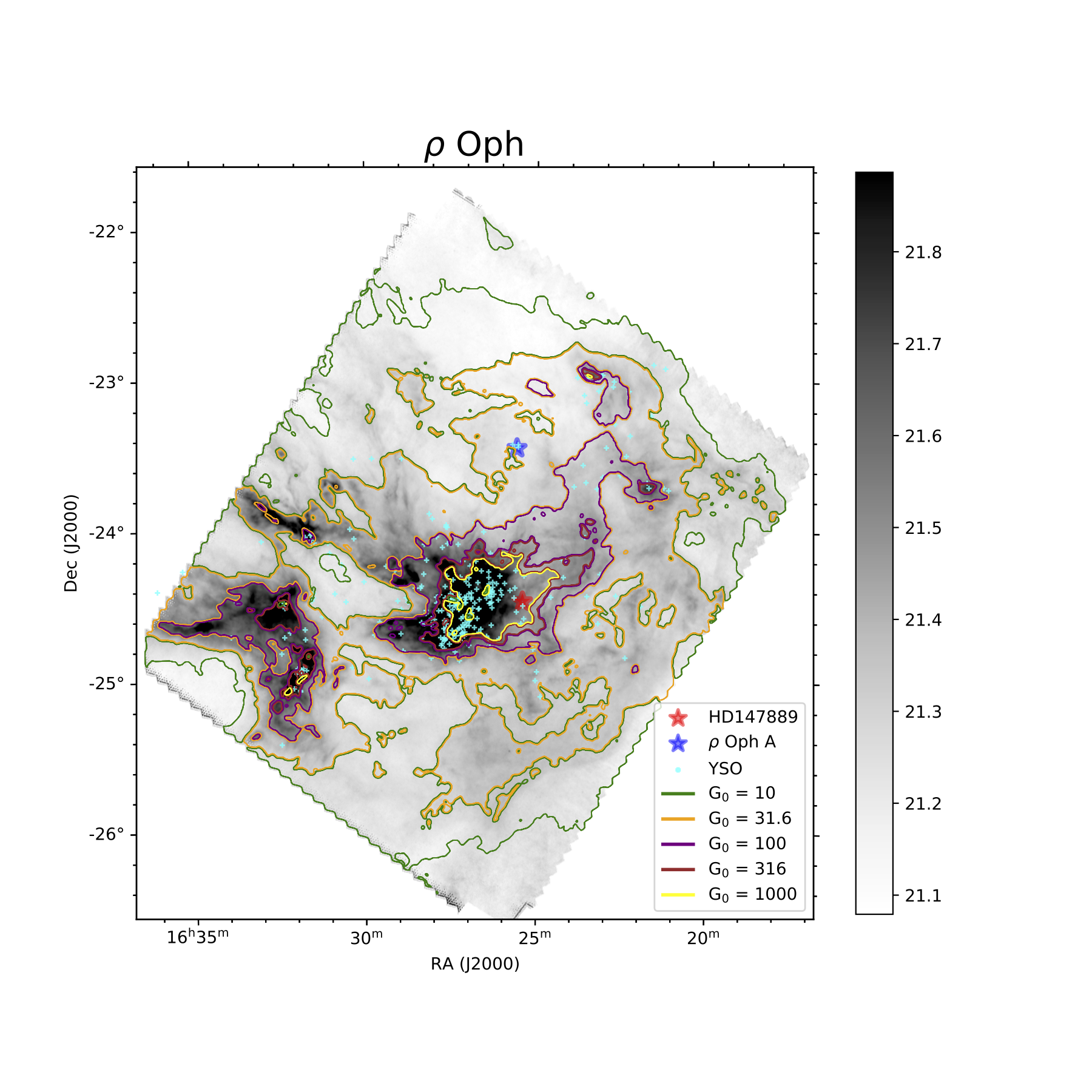} 
 \caption{Contours of UV intensity overlaid on H$_2$ column density map of the $\rho$ Oph region. The labels of the figure are the same as that in Fig. \ref{fig:l1172}.}
 \label{fig:rho Oph}
 \end{figure}

The UV radiation distribution of $\rho$ Oph cloud is shown in Fig. \ref{fig:rho Oph}. A strong correlation between UV intensity and star distribution was found. The UV intensity G$_0$ can exceed 1000 in dense gas region and regions around OB stars.

\subsection{Perseus}

\begin{figure}[h!]
\centering
\includegraphics[width=0.80\textwidth,height=0.70\textwidth,trim={{0.01\textwidth} {0.60\textwidth}  {0.05\textwidth}  {0.1\textwidth} },clip]{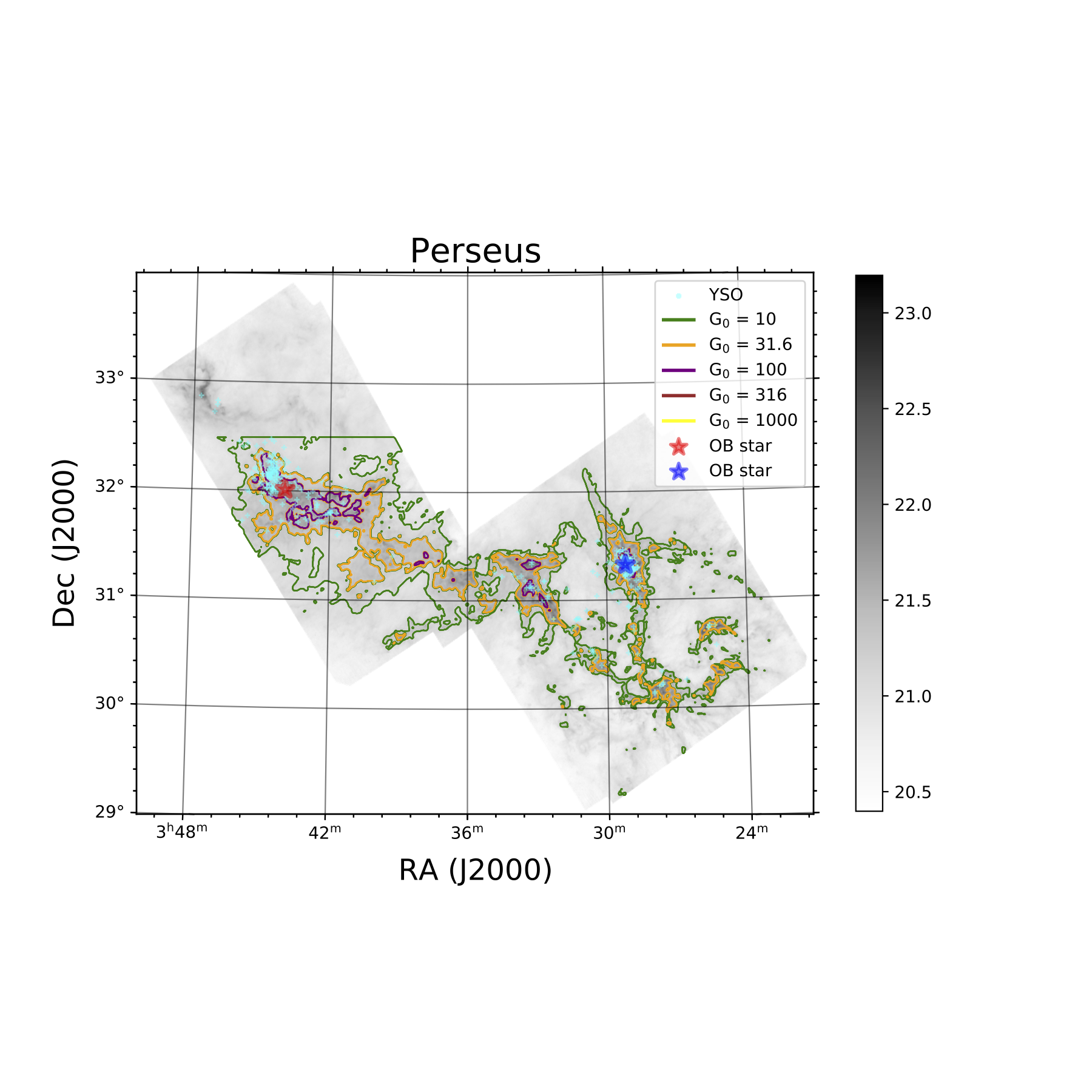}
\caption{Contours of UV intensity overlaid on H$_2$ column density map of the Perseus region. The labels of the figure are the same as that in Fig. \ref{fig:l1172}.}
\label{fig:perseus}
\end{figure}

The Perseus molecular clouds is $\sim$ 250 pc away with sky coverage of $\sim$ 10 deg{$^2$}. It is a low and intermediate-mass star-forming region. The total mass of the Perseus molecular cloud is about 6586 M$_{\odot}$, one third of which is dense gas \citep{2014ApJ...782..114E}. As shown in Fig. \ref{fig:perseus}, UV radiation field correlates with the locations of OB stars and YSOs in this region.

\subsection{Pipe}
\begin{figure}[h!]
 \centering
 \includegraphics[width=0.80\textwidth,height=0.70\textwidth,trim={{0.10\textwidth} {0.80\textwidth}  {0.05\textwidth}  {0.1\textwidth} },clip]{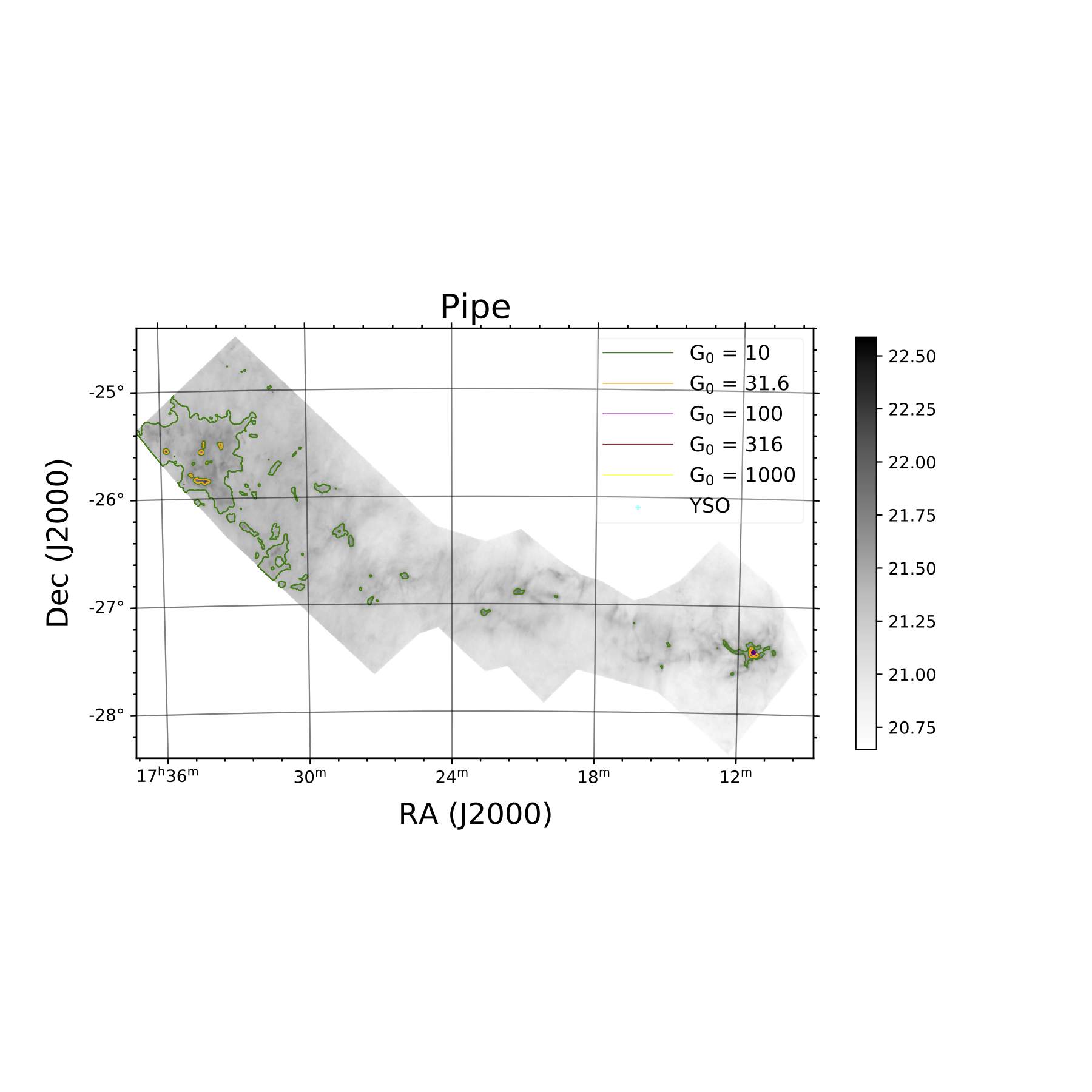}
 \caption{Contours of UV intensity overlaid on H$_2$ column density map of the Pipe region. The labels of the figure are the same as that in Fig. \ref{fig:aquila}.}
 \label{fig:pipe}
 \end{figure}

The Pipe Nebula has a distance of $\sim$ 145 pc \citep{2007A&A...470..597A}. Composed of an elongated dark cloud with length of 18 pc, the Pipe Nebula is one of the closest  star-forming regions. The Pipe Nebula is an ideal target for investigating core formation. The mass of the Pipe nebula is about 1.7 $\times$ 10$^4$ M$_{\odot}$ \citep{ 2006A&A...454..781L}. A few identified YSOs were found in this region\citep{2012A&A...541A..63P}. As shown in Fig.\ref{fig:pipe}, the UV intensity is very low ($G_0<$ 31.6) toward most region of this molecular complex.

\subsection{Serpens}

\begin{figure}[h!]
\centering
\includegraphics[width=0.80\textwidth,height=0.70\textwidth,trim={{0.01\textwidth} {0.55\textwidth}  {0.05\textwidth}  {0.1\textwidth} },clip]{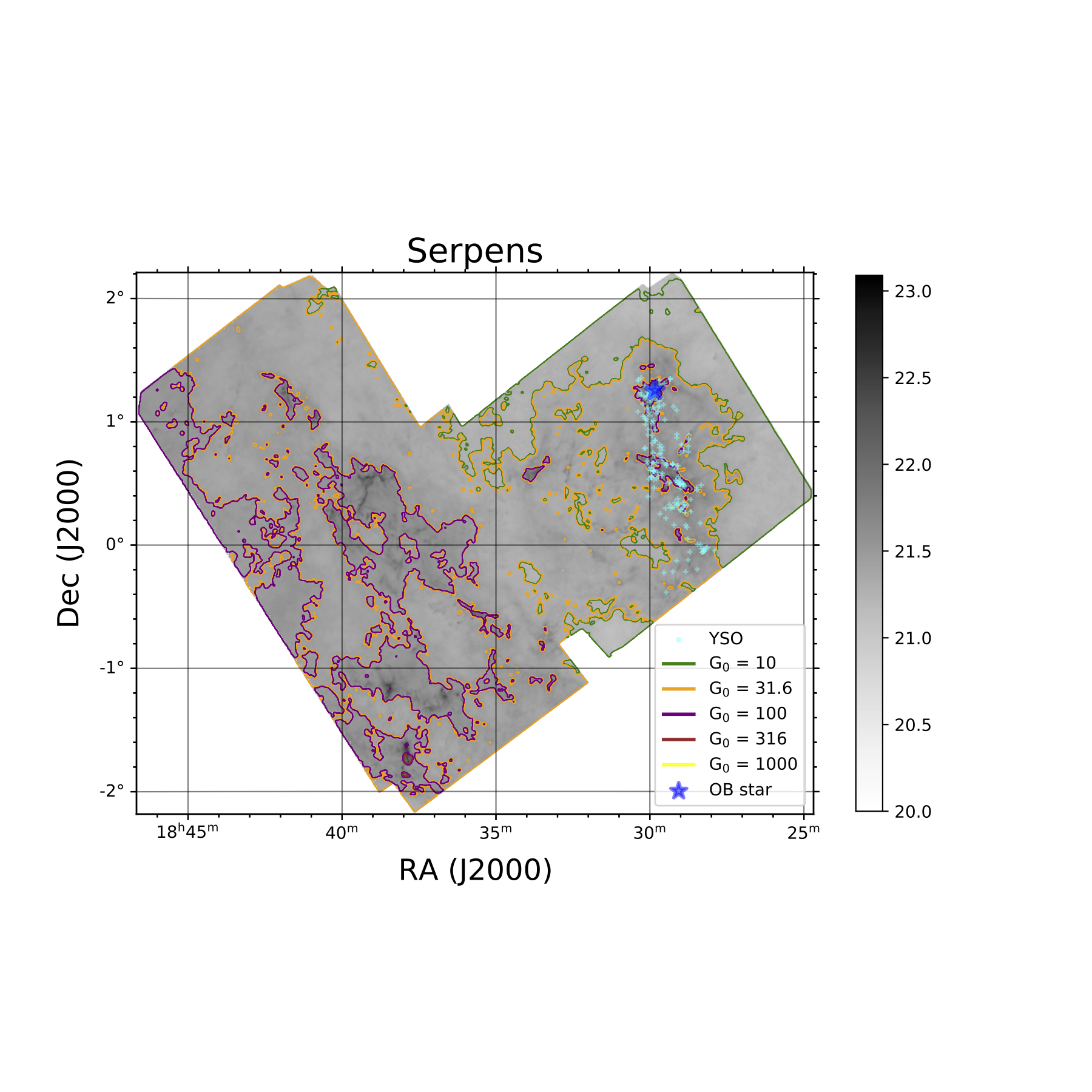}
\caption{Contours of UV intensity overlaid on H$_2$ column density map of the Serpens region. The labels of the figure are the same as that in Fig. \ref{fig:l1172}}

\label{fig:serpens}
\end{figure}

The Serpens star-forming region located at $\sim$ 429 pc was covered in $\sim$ 15 deg{$^2$} by HGBS \citep{2014ApJ...782..114E}. Its total mass is $\sim$ 6583 M$_{\odot}$, two thirds of which is dense gas \citep{2014ApJ...782..114E}. About 81 percent of the prestellar cores are found in the filamentary structure of Serpens \citep{2021MNRAS.500.4257F}. 
Serpens is confirmed to be a low-mass and active star-forming region at a young age. As shown in Fig. \ref{fig:serpens}, lots of YSOs and an OB star were found in this region.

\subsection{Taurus}
\begin{figure}[h!]
 \centering
 \includegraphics[width=0.80\textwidth,height=0.70\textwidth,trim={{0.10\textwidth} {0.50\textwidth}  {0.05\textwidth}  {0.1\textwidth} },clip]{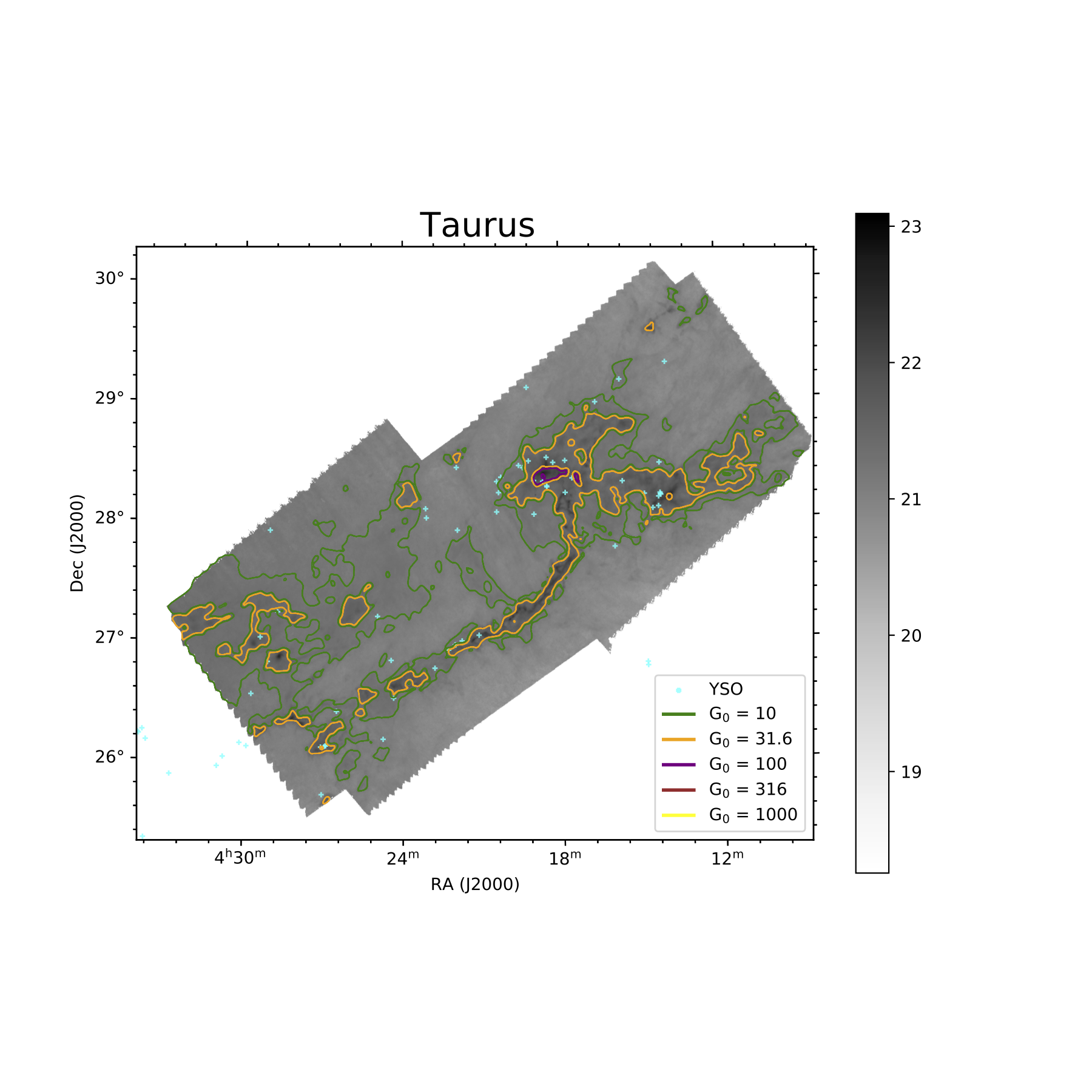}
 \caption{Contours of UV intensity overlaid on H$_2$ column density map of the Taurus region. The labels of the figure are the same as that in Fig. \ref{fig:aquila}.}
 \label{fig:taurus}
 \end{figure}

The distance of Taurus cloud to our solar system is about 140\,pc \citep{2013A&A...550A..38P}. The total mass of Taurus molecular cloud is about 2-4 $\times$ 10$^4$ M$_{\odot}$, ten percent of which is dense gas \citep{2014ApJ...782..114E}. HGBS covered about 52 deg$^2$ of this region \citep{2013MNRAS.432.1424K}. No OB stars was found in the Taurus molecular cloud. As shown in Fig. \ref{fig:taurus}, the UV radiation  intensity correlates with the distribution of dense gas. 

\end{document}